\documentclass{article}

\usepackage{arxiv}

\usepackage[utf8]{inputenc}
\usepackage[T1]{fontenc}
\usepackage{amsmath,amssymb}
\usepackage{graphicx}
\usepackage{booktabs}
\usepackage{subcaption}
\usepackage{microtype}
\usepackage{url}
\usepackage[numbers,square,comma]{natbib}
\usepackage[dvipsnames]{xcolor}
\definecolor{AccentBlue}{RGB}{20,55,100}
\usepackage[colorlinks=true,
            linkcolor=AccentBlue,
            citecolor=AccentBlue,
            urlcolor=AccentBlue]{hyperref}
\usepackage[capitalize,noabbrev]{cleveref}
\usepackage{doi}
\usepackage{titlesec}

\makeatletter
\renewcommand{\normalsize}{%
  \@setfontsize\normalsize\@xpt{12.5}%
  \abovedisplayskip      7\p@ \@plus 2\p@ \@minus 5\p@
  \abovedisplayshortskip \z@ \@plus 3\p@
  \belowdisplayskip      \abovedisplayskip
  \belowdisplayshortskip 4\p@ \@plus 3\p@ \@minus 3\p@
}
\normalsize
\renewcommand{\@toptitlebar}{%
  \hrule height 0.4pt%
  \vskip 0.25in%
  \vskip -\parskip%
}
\renewcommand{\@bottomtitlebar}{%
  \vskip 0.29in%
  \vskip -\parskip%
  \hrule height 0.4pt%
  \vskip 0.09in%
}
\makeatother

\titleformat{\section}
  {\large\scshape\bfseries\color{AccentBlue}}
  {\thesection.}{0.5em}{}
  [{\color{AccentBlue!40}\titlerule[0.4pt]}]
\titleformat{\subsection}
  {\normalsize\bfseries\itshape}
  {\thesubsection.}{0.5em}{}
\titleformat{\subsubsection}
  {\normalsize\bfseries}
  {\thesubsubsection.}{0.5em}{}
\titleformat{\paragraph}[runin]
  {\normalsize\bfseries}
  {}{0em}{}[]
\titlespacing*{\section}{0pt}{14pt plus 4pt minus 2pt}{7pt plus 1pt}
\titlespacing*{\subsection}{0pt}{10pt plus 2pt minus 1pt}{3pt plus 1pt}
\titlespacing*{\subsubsection}{0pt}{8pt plus 2pt minus 1pt}{2pt}
\titlespacing*{\paragraph}{0pt}{6pt plus 2pt minus 1pt}{1em}

\renewenvironment{abstract}{%
  \begin{center}{\large\bfseries\scshape\color{AccentBlue}Abstract}\end{center}%
  \noindent\textcolor{AccentBlue!60}{\rule{\linewidth}{0.5pt}}\vspace{5pt}%
  \begin{quote}
}{%
  \end{quote}%
  \vspace{-2pt}%
  \noindent\textcolor{AccentBlue!60}{\rule{\linewidth}{0.5pt}}%
}

\graphicspath{{./}{figures/}}

\newcommand{\bdiff}{\beta_{\mathrm{diff}}}
\newcommand{\Ntarget}{N_{\mathrm{target}}}
\newcommand{\Cn}{C_n}
\newcommand{\Dn}{D_n}
\newcommand{\eps}{\varepsilon}
\newcommand{\Gsize}{\lvert G\rvert}
\newcommand{\ci}[2]{[#1,\,#2]}

\title{Measuring the Symmetry--Data Exchange Rate\\[2pt]
\large A Controlled Measurement Under Exactly Known Symmetry}

\author{%
  \href{https://orcid.org/0009-0004-4344-141X}{\includegraphics[scale=0.06]{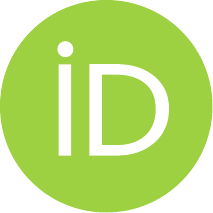}\hspace{1mm}Ahmed M. Adly} \\
  Independent Researcher \\
  Egypt \\
}

\renewcommand{\shorttitle}{Measuring the Symmetry--Data Exchange Rate}

\hypersetup{
  pdftitle={Measuring the Symmetry-Data Exchange Rate},
  pdfauthor={Ahmed M. Adly},
  pdfsubject={cs.LG},
  pdfkeywords={equivariance, sample complexity, inductive bias, data efficiency, exchange rate, pre-specified controls}
}

\date{}

\begin{document}
\maketitle

\begin{abstract}
Equivariance theory predicts that an architectural symmetry prior reduces sample complexity by a factor of $\Gsize$; this is widely cited but rarely measured as a scaling law with controls that separate the prior from its confounds. On a controlled $\Cn$-symmetric task, we report three findings. First, a \emph{wrong-group} control with identical orbit size and matched compute is \emph{worse} than no constraint (joint pairwise CI $\ci{+0.79}{+3.26}$ excludes zero, robust across estimators); misaligned constraint is actively harmful, not merely unhelpful. Second, an augmentation baseline equipped with \emph{test-time} orbit averaging matches the equivariant model exactly --- bit-identical per-epoch validation curves across matched cells --- so the architecture-vs-augmentation gap is conditional on \emph{asymmetric} test-time computation, not unconditional. Third, the relative exchange rate $\bdiff = 1.28$ is consistent in sign and order of magnitude with the theoretical $1.0$ (single-level CI $\ci{+0.92}{+2.05}$); the more conservative two-level bootstrap (seeds $\times$ group sizes) widens this to $\ci{-0.63}{+1.72}$, \emph{including zero}, and a finer-$N$ replication on a $\sqrt{2}$-spaced grid is inconclusive (point estimate $-0.82$). The methodological contributions --- the relative-rate estimator that cancels the shared-difficulty confound, the wrong-group control, and a pre-specified failure taxonomy --- transfer to any inductive bias whose strength can be parameterised. \emph{Honest scoping}: the primary estimator $\bdiff$ was adopted post-hoc after the initial analysis revealed a positive-slope identifiability problem; the design was never externally pre-registered; and the headline number rests on an OLS slope over seven group sizes on a coarse $N$ grid. This is an exploratory study, not a confirmatory measurement; the wrong-group result is the cleanest finding and the one we report with the most confidence. A registered replication on fresh seeds is future work.
\end{abstract}


\section{Introduction}
\label{sec:intro}

A model that is built to respect a symmetry does not have to learn that symmetry from data. If a task's labels are invariant under a finite group $G$ acting on the inputs, an architecture that is invariant by construction~\citep{cohenwelling2016} collapses each $\Gsize$-element orbit to a single effective example, and analyses of learning under invariance predict that the sample complexity falls by a factor of about $\Gsize$~\citep{mei2021,bietti2021}. Equivariance has driven sample-efficiency gains across molecules~\citep{batzner2022,anderson2019}, point clouds~\citep{thomas2018}, and sets and graphs~\citep{zaheer2017,defferrard2016}, and the geometric-deep-learning program has organized these gains around the group acting on the domain~\citep{bronstein2021}. Data is the binding constraint in many of the regimes where this matters most --- small scientific corpora, expensive labels, low-data deployment --- which is precisely where a $\Gsize$-fold reduction would be decisive.

Yet the $\Gsize$ prediction is more often cited than measured. Empirical studies typically fix the sample size and report an accuracy gap, or fix the accuracy and report an efficiency ratio at a single operating point; they rarely trace the \emph{sample-complexity scaling law} --- how the samples needed to reach a target accuracy vary with $\Gsize$ --- and they rarely include controls that isolate the structural prior from the confounds that travel with it. Weight sharing reduces effective capacity; orbit pooling can be read as implicit data augmentation; any architectural constraint regularizes. A measurement that does not separate these cannot distinguish ``the correct symmetry helps'' from ``some constraint of this strength helps.'' The gap we address is therefore methodological: how to turn ``structure helps'' into a calibrated, falsifiable exchange rate.

We build a task in which the symmetry group is known exactly and its order $n$ is a free knob, fix five model families before collecting data --- the correct equivariant model and four controls, each constructed to rule out one specific alternative explanation --- and pre-specify the statistical analysis and a failure taxonomy that states in advance what would count as evidence \emph{against} the hypothesis. We then measure the exchange rate. Our central claim, stated honestly: \emph{on this controlled task, with the relative-rate estimator adopted post hoc, the equivariant model's sample-complexity slope sits below the unconstrained model's by a margin consistent with the theoretical one-bit-per-bit prediction in sign and order of magnitude --- $\bdiff = 1.28$ (single-level percentile CI $\ci{0.92}{2.05}$; two-level seeds-$\times$-group-sizes CI $\ci{-0.63}{+1.72}$ which includes zero) --- and a wrong-group control of equal orbit size is harmful rather than helpful (joint pairwise CI excludes zero).} The headline number is not a precise measurement; the qualitative direction is robust to every check we run.

\paragraph{Contributions.}
\begin{enumerate}
\item \textbf{A confound-free measurement methodology.} We introduce a \emph{relative} exchange-rate estimator that cancels the shared task-difficulty scaling which otherwise makes the naive prediction untestable, together with a joint pairwise bootstrap and a pre-specified failure taxonomy (\cref{sec:methods}). The methodology transfers to any inductive bias whose strength can be parameterized.
\item \textbf{The estimated rate.} On a $\Cn$-symmetric task the point estimate is $\bdiff = 1.28$ (two-level seeds-$\times$-group-sizes CI $\ci{-0.63}{+1.72}$ includes zero; single-level percentile CI $\ci{+0.92}{+2.05}$ excludes zero); the rate retains $\geq 88\%$ of its central value across the $\varepsilon$ sweep through $30\%$ label corruption (\cref{sec:results}). We report the qualitative direction, not the precise numerical value, as the contribution.
\item \textbf{Controls that isolate alignment from constraint.} A wrong-group control of identical orbit size is \emph{worse} than no constraint, ruling out generic constraint. Training-time-only orbit augmentation (single-input forward pass at test time) cannot reach the target accuracy where the equivariant model succeeds; a CPU replication shows that \emph{the same augmented model with test-time orbit averaging matches the equivariant model exactly} (\cref{sec:augmentation,app:cpu_replication}). The architecture-vs-augmentation gap is therefore specifically about training-time-only augmentation under asymmetric test-time computation, not about augmentation in general.
\item \textbf{A fully reproducible artifact.} The complete experiment runs in roughly $90$ minutes on one GPU; all code, the design document, the configuration hash, and every per-run record are released. Pre-specified measurements that failed calibration are reported as negative results (\cref{app:estimators}), not omitted.
\end{enumerate}

We are equally explicit about what the result is. We report an \emph{operational} exchange rate --- an empirical estimate conditional on a specified target accuracy and scaling model, in bit-equivalent ($\log_2$) units --- not a universal information-theoretic equivalence; its value is a measurement in one setting, not a constant we claim holds elsewhere. The saving is in data, not compute (it leaves total training FLOPs to the target roughly fixed); the task is $2$-D and synthetic with exact symmetry; and the study is exploratory --- the relative-rate estimator was adopted after an initial analysis, so the confirmatory weight rests on a registered replication, which is future work.

\section{Related work}
\label{sec:related}

\paragraph{Inductive bias and sample complexity, broadly.} The general claim that prior structure reduces sample complexity goes back at least to~\citet{baxter2000}'s formal model of inductive-bias learning, where bias is treated as a hypothesis-space restriction that improves the bias-variance trade-off. \citet{wolpert1996} provides the complementary negative result: without a prior aligned to the task, no learning algorithm dominates uniformly. The factor-$\Gsize$ claim for group equivariance is a specific quantitative instance of the general principle.

\paragraph{Equivariant architectures and the $\Gsize$ claim.} Group-equivariant convolutional networks formalized weight sharing over a group and articulated the expectation that equivariance improves sample efficiency~\citep{cohenwelling2016}; subsequent work generalized the construction to compact groups~\citep{kondortrivedi2018} and to steerable representations of Euclidean groups~\citep{weilercesa2019}, and the broader program frames learning through the symmetry acting on the domain~\citep{bronstein2021}. Our equivariant model is a textbook regular-representation network in this lineage; we claim no architectural novelty, only a controlled measurement of its sample-complexity scaling.

\paragraph{Theory of sample complexity under invariance.} The factor-$\Gsize$ intuition is made precise in analyses of invariant kernels and random features~\citep{mei2021,bietti2021}, which quantify how invariance shrinks the effective hypothesis space. These results predict the \emph{form} of the scaling we measure; we supply the controlled empirical curve. We note an important caveat: these analyses are derived for invariant kernels and random feature models, not for finite-width ReLU MLPs trained by Adam, and whether the quantitative $\Gsize$-fold prediction transfers to our model class is an empirical question, not an analytical fact. We therefore benchmark our measurement against the theoretical prediction without claiming a formal bridge between the two settings, and our agreement claim is in sign and order of magnitude, not in precise numerical confirmation of a theorem that technically applies to a different model class.

\paragraph{Augmentation versus architecture.} Whether data augmentation can substitute for built-in invariance is a long-standing question. \citet{elesedyzaidi2021} give a theorem separating equivariant models from their augmented counterparts: the former enjoy a strict generalization benefit, so the two mechanisms are not equivalent. Our augmentation result is the empirical complement, conditional on a specific evaluation design: the augmented model in our experiment uses a single-input forward pass at test time, while the equivariant model effectively averages over the orbit through its pooling layer, so the comparison is between \emph{architectural equivariance with orbit pooling at inference} and \emph{training-time augmentation only}; on this comparison, augmentation does not merely lose efficiency, it fails to reach the target accuracy at all in the regime where the equivariant model succeeds. A test-time-averaged augmentation variant is the natural counterfactual; we ran it in a CPU replication and found that it matches the equivariant model exactly at every $n$ (\cref{sec:augmentation,app:cpu_replication}). The architecture-vs-augmentation gap is therefore conditional on asymmetric test-time computation, not unconditional.

\paragraph{Methodology.} Our inference rests on the non-parametric bootstrap, including the paired-difference variant~\citep{efron1993}; our insistence on specifying the analysis and the failure modes before interpreting results follows the argument that, in fields without registration, undisclosed analytic flexibility silently inflates apparent findings~\citep{gelmanloken2014}.

\paragraph{The gap, stated against the literature.} Most prior empirical work compares an equivariant model to a single unconstrained baseline and reports an accuracy gap at fixed sample size. We instead trace the slope of sample complexity against $\Gsize$, with an orbit-size-matched wrong-group control, a capacity- and regularization-matched baseline, and a relative-rate estimator that removes the task-difficulty confound. To make the gap concrete: \citet{cohenwelling2016} report rotation-equivariant CNN accuracy gains on rotated MNIST and CIFAR at fixed training-set sizes, but do not trace a sample-complexity scaling law against $\Gsize$ and do not include an orbit-size-matched misaligned-symmetry control. \citet{weilercesa2019} run systematic comparisons across the $E(2)$ subgroup lattice and report data-efficiency improvements, but again at fixed operating points rather than as a fitted slope, and without a misaligned-symmetry baseline. \citet{batzner2022} demonstrate striking data efficiency for $E(3)$-equivariant interatomic potentials on quantum-chemistry datasets and report learning curves at multiple training sizes, but the $\Gsize$ knob is the architectural choice itself rather than a continuously parameterised group order, and the comparison is to non-equivariant baselines rather than to a same-orbit misaligned control. None of these isolate the scaling exponent from the task-difficulty confound, and none separate ``alignment'' from ``constraint'' through a matched-orbit wrong-group control. To our knowledge the combination --- a fitted scaling law with adversarial controls and a pre-specified taxonomy --- has not been reported.

\section{Methodology}
\label{sec:methods}

\subsection{Task}
\label{sec:task}
\begin{figure}[t]
\centering
\begin{subfigure}{0.66\linewidth}
\centering
\includegraphics[width=\linewidth]{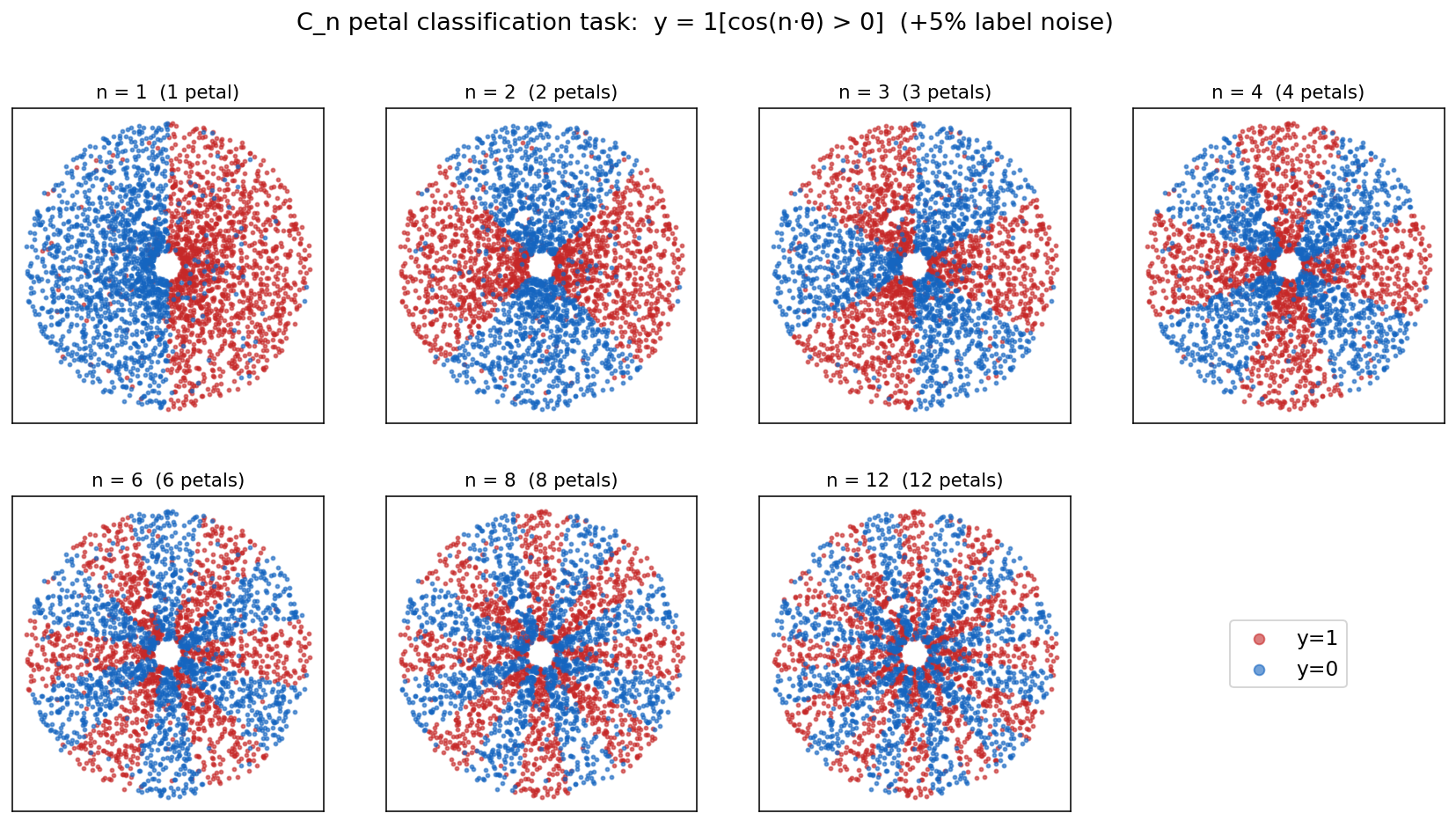}
\caption{The $\Cn$-petal task across group orders.}
\label{fig:task-gallery}
\end{subfigure}
\hfill
\begin{subfigure}{0.31\linewidth}
\centering
\includegraphics[width=\linewidth]{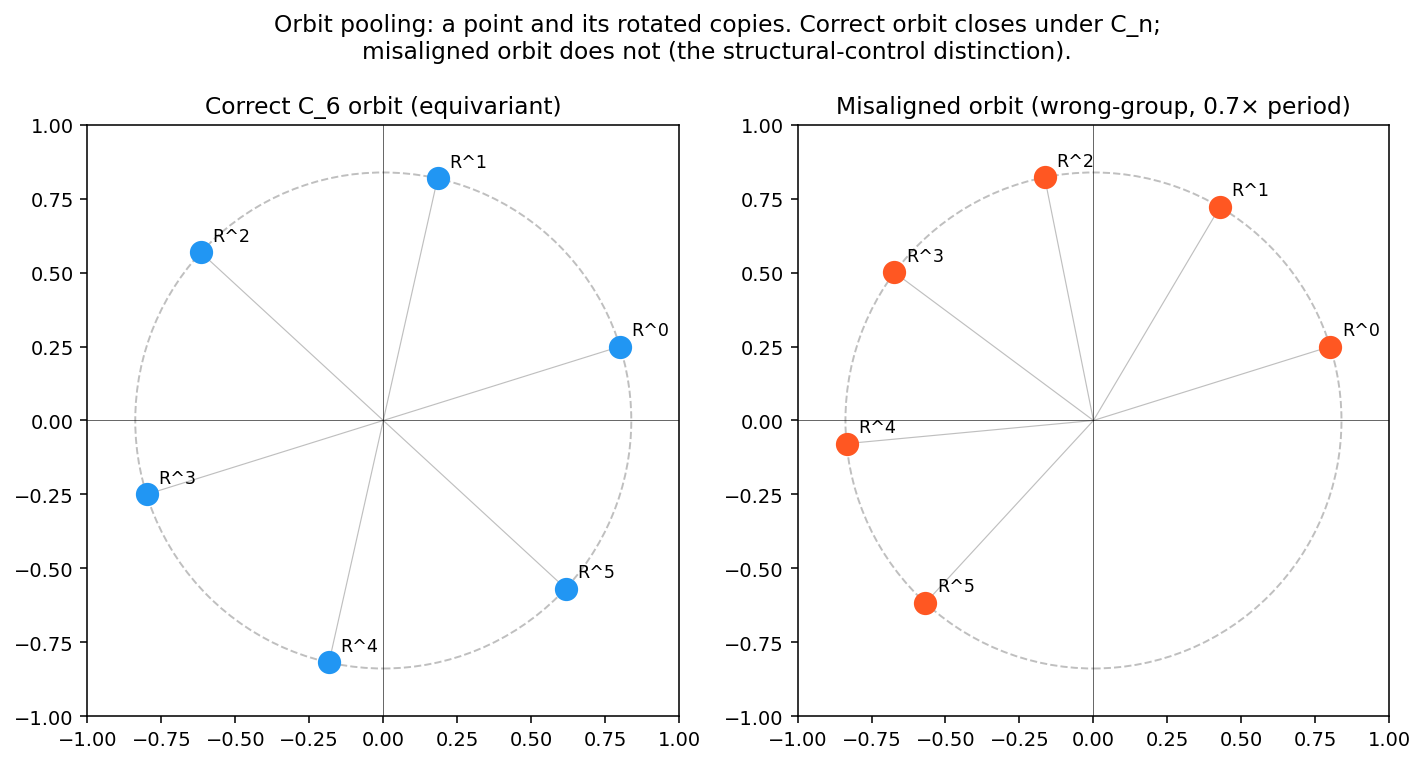}
\caption{Correct vs.\ misaligned orbit.}
\label{fig:task-orbits}
\end{subfigure}
\caption{\textbf{The task and the control idea.} (\subref{fig:task-gallery}) Inputs are sampled on an annulus and labeled $\mathbf{1}[\cos(n\theta) > 0]$ for $n \in \{1,2,3,4,6,8,12\}$, an alternating angular pattern whose symmetry group is exactly $\Cn$ by construction; the number of positive lobes is $n$, while the full alternating pattern has $2n$ angular sectors. (\subref{fig:task-orbits}) The equivariant model averages a point over its correct $\Cn$ orbit; the wrong-group control averages over an orbit of \emph{equal size} at a misaligned period. \emph{Takeaway:} the symmetry is known exactly, and the wrong-group control differs from the treatment only in orbit alignment --- isolating alignment from constraint.}
\label{fig:task}
\end{figure}

Inputs $x \in \mathbb{R}^2$ are sampled uniformly from an annulus with radius $r \in [0.1, 1.0]$ and angle $\theta \in [0, 2\pi)$. The clean label is
\begin{equation}
y_{\mathrm{clean}}(x) = \mathbf{1}\!\left[\cos(n\,\theta(x)) > 0\right],
\end{equation}
an alternating pattern of angular ``petals.'' The clean label is invariant under rotation by any multiple of $2\pi/n$, so its symmetry group is the cyclic group $\Cn$. The label is in fact also reflection-invariant, so the full symmetry group is the dihedral $\Dn$; our equivariant model exploits only the rotation subgroup $\Cn$, which makes any measured advantage a conservative lower bound on what a $\Dn$-equivariant model would obtain (\cref{sec:discussion}). A fixed $5\%$ Bernoulli label-noise rate is applied to every split. The group order $n \in \{1,2,3,4,6,8,12\}$ is a free integer knob (\cref{fig:task}).

\subsection{Why this task is diagnostic.}
\label{sec:diagnostic}
The task is chosen to satisfy, simultaneously, four properties that real datasets rarely combine: (i) the true symmetry group is \emph{known analytically}, not estimated; (ii) $\Gsize = n$ is a free parameter, so a scaling law can be \emph{designed} rather than inferred; (iii) the problem is non-trivial in absolute terms --- at $n = 12$ a model must learn twelve alternating $30^{\circ}$ regions that generalize to unseen angles; and (iv) it admits a natural family of \emph{misaligned} controls (the same orbit-pooling architecture at the wrong rotation angle). Adversarial checks confirm the task has no shortcut: a logistic-regression probe on raw coordinates reaches at most $0.55$ accuracy, classes are balanced to within $1\%$ of $0.5$, train and validation sets share no coordinate, and the validation labels are $\Cn$-invariant to within rounding (\cref{app:tables}).

\subsection{Model families}
\label{sec:models}
All five families share the identical fully-connected architecture (hidden width $32$, two hidden ReLU layers, single output logit; $2{\cdot}32 + 32 + 32{\cdot}32 + 32 + 32 + 1 = 1{,}185$ trainable parameters per network), so capacity is matched exactly --- not approximately. The control families differ from the equivariant treatment only in how data flow through this shared architecture (orbit pooling, augmentation, regularization, or none). Each control is constructed to rule out one specific alternative to the structural-prior explanation.
\begin{itemize}
\item \textbf{Equivariant} (treatment). For input $x$ the network forms all $n$ rotated copies $R_k x$, applies a shared multi-layer perceptron to each, and averages the scalar outputs; this composition is $\Cn$-invariant by construction (verified by unit test). Orbit pooling itself has no learnable parameters.
\item \textbf{Wrong-group} (rules out ``any orbit pooling helps''). The identical orbit-pooling architecture, but with rotation angles misaligned by an irrational factor of $0.7\times$ the correct period, chosen so the orbit accidentally aligns with no $C_k$ subgroup. Same compute, same orbit size, same effective degrees of freedom --- only the alignment differs.
\item \textbf{Augmented} (rules out ``it is just more data coverage''). An unconstrained network that during training sees all $n$ rotated copies and averages their logits in the loss, but at evaluation uses the standard single-input forward pass. It receives the same orbit information as the equivariant model but is not constrained to respect it. We do not use test-time orbit averaging here --- a distinct intervention from training-time augmentation --- in order to hold test-time computation equal across all families. We separately ran a test-time-averaged variant (\texttt{augmented\_tta}) in a CPU replication; it matches the equivariant model exactly at every $n$, with bit-identical per-epoch validation curves across all matched cells (\cref{app:cpu_replication}).
\item \textbf{Vanilla} (rules out raw capacity). A standard network with parameter count $\geq$ the equivariant model's.
\item \textbf{Regularized} (rules out effective-DOF reduction). The vanilla network with $L_2$ weight decay $\lambda = 10^{-3}$, set by a pilot to remove roughly the degrees of freedom that orbit averaging removes.
\end{itemize}

\subsection{Symmetry breaking.}
\label{sec:epsilon}
To test robustness to a violated symmetry we corrupt a fraction $\eps$ of the \emph{training} labels (before noise) by replacing them with a $C_1$-asymmetric reference $\mathbf{1}[\cos(\theta - \alpha) > 0]$ for a uniformly sampled $\alpha$; validation and test labels are the clean $\Cn$ labels in every condition. We sweep $\eps \in \{0, 0.1, 0.2, 0.3\}$. Note that $\eps$ corrupts the training \emph{signal}; the underlying task geometry remains exactly $\Cn$-symmetric. This is a weaker perturbation than corrupting the task geometry itself: the $\varepsilon$ sweep tests robustness to label corruption, not to tasks whose geometric structure is only approximately $\Cn$-symmetric; the latter is a distinct and open question (\cref{sec:discussion}).

\subsection{Sample-complexity metric and the naive law.}
\label{sec:metric}
We train each model with Adam (learning rate $10^{-3}$, batch size $64$) for up to $500$ epochs and define
\begin{equation}
\Ntarget = \min\{N : \text{at least } 3 \text{ of } 5 \text{ seeds reach validation accuracy} \geq 0.80\},
\end{equation}
over the training-set grid $N \in \{50, 100, 200, 400, 800, 1600, 3200, 6400\}$. The target $T = 0.80$ was pre-specified; we had conjectured the rate would be non-decreasing in $T$ within $[0.70, 0.85]$, but the CPU replication of \cref{app:cpu_replication} \emph{refutes} this conjecture --- $\bdiff$ in fact decreases from $\approx +0.96$ at $T = 0.70$ to $\approx +0.14$ at $T = 0.85$. The metric assumes that the $3$-of-$5$ pass-rate is non-decreasing in $N$ at each (model, $n$) cell; we verified this empirically on the CPU replication (\cref{app:cpu_replication}): zero of $42$ (model, $n$) cells exhibited a majority-vote monotonicity violation. (Eight of $210$ per-seed cells exhibited isolated non-monotonicity at the single-seed level, none of which crossed the majority-vote threshold in a direction that changed $\Ntarget$.) Violations at the majority-vote level would bias the slope estimate unpredictably; we observed none. The naive prediction is a scaling law $\log_2 \Ntarget(n) = \beta_0 + \beta_1 \log_2(n)$ with slope $\beta_1 \approx -1$ for the equivariant model (orbit averaging halves the requirement when $n$ doubles) and $\beta_1 \approx 0$ for the baseline. This hides an assumption: that the baseline's requirement is constant in $n$. It is not. The petal task gets harder as $n$ grows --- the decision boundary alternates more finely --- so \emph{every} model's absolute slope is positive (vanilla $+1.90$, $\ci{+1.06}{+3.49}$; equivariant $+0.62$, $\ci{-0.24}{+1.69}$; \cref{sec:results}). The naive prediction is therefore not identifiable from absolute slopes.

\subsection{The relative exchange rate.}
\label{sec:bdiff}
The quantity that the theory actually predicts, and that is identifiable, is the \emph{relative} exchange rate
\begin{equation}
\bdiff(\text{treatment}) \;:=\; \text{slope of } \log_2\!\left(\frac{\Ntarget^{\text{vanilla}}}{\Ntarget^{\text{treatment}}}\right) \text{ against } \log_2(n) \;=\; \beta_1(\text{vanilla}) - \beta_1(\text{treatment}).
\end{equation}
If the treatment needs $N_{\text{t}} = c\,n^{\alpha}\,n^{-1}$ samples and the baseline needs $N_{\text{v}} = c\,n^{\alpha}$ for a shared task-difficulty exponent $\alpha$, then $\log_2(N_{\text{v}}/N_{\text{t}}) = \log_2 n$, whose slope against $\log_2 n$ is exactly $+1$. The relative rate cancels the shared $\alpha$ and recovers the original theoretical target. We disclose that adopting $\bdiff$ as the primary estimator was a \emph{post-hoc} decision: the pre-specified analysis targeted the absolute slope $\beta_1 \approx -1$, and we moved to the relative rate after an initial analysis revealed the positive-slope confound (\cref{sec:status}). The relative rate is a re-expression of the same theoretical content on the same data, not a search over estimators, but because the choice followed data inspection the present result is exploratory and the confirmatory weight rests on the registered replication. To clarify the mechanism: the relative estimator was adopted because the $|G|$-fold theoretical prediction translates directly into a relative slope (the log-ratio cancels the shared difficulty term $\alpha$ by construction), not because it yielded a significant or favourable result; had the relative slope been negative or indistinguishable from zero, we would report it as such --- and the wrong-group relative rate of $-0.77$ confirms that the estimator is informative in both directions. Nonetheless, we cannot fully rule out that the decision to log-difference slopes, rather than another transformation, was partially influenced by the direction of the data, which is why the exploratory designation stands.

\paragraph{Identifiability and the meaning of ``bits.''} The relative rate is identifiable under one explicit assumption: that the baseline and the treatment share the same task-difficulty exponent $\alpha$, so that it cancels in the log-ratio. This is testable rather than assumed away --- if the two models scaled with different $\alpha$, the log-ratio would not be linear in $\log_2 n$, which would appear as curvature in \cref{fig:scaling}; we do not observe such curvature. A subtler threat is convergence heterogeneity: if one model converges more reliably at a given $N$ independent of function-class effects, this could produce slope differences without truly different difficulty exponents. The wrong-group control is partially diagnostic here: it uses identical orbit-pooling mechanics, so convergence benefits from orbit pooling alone would show up as a positive wrong-group rate; instead it is $-0.77$, suggesting the rate captures structural alignment rather than convergence behaviour alone. Where the assumption fails --- for instance if a constraint changes not just the offset of the sample-complexity curve but the rate at which the task hardens with $\Gsize$ --- the slope of the log-ratio is no longer a pure exchange rate and must be read as a descriptive summary of relative efficiency. We use ``bits'' only as a name for $\log_2$ units: quotienting an orbit of size $n$ removes $\log_2 n$ bits of redundancy from the function class, and the rate reports how many bits of data that removal buys. Throughout, $\bdiff$ is an \emph{operational} quantity --- the slope of a fitted sample-complexity ratio under a chosen target accuracy and scaling model. Here ``one bit'' means precisely a unit step in $\log_2 \Gsize$ (a doubling of the group order), so the rate is the empirical slope of $\log_2(\Ntarget^{\text{vanilla}}/\Ntarget^{\text{equivariant}})$ against $\log_2 \Gsize$ --- equivalently, the orbit redundancy removed from the function class. We do \emph{not} use ``bit'' in a coding-theoretic or compression sense, and we intend no claim that the numerical value is universal across tasks, groups, or architectures. \Cref{sec:robustness} characterizes the estimate's sensitivity to these choices.

\paragraph{Structural floor from grid discretization.} The discrete $N$ grid (factors of $2$ apart) imposes a quantization on $\Ntarget$: the per-$n$ log-ratio $\log_2(\Ntarget^{\text{vanilla}}/\Ntarget^{\text{treatment}})$ is forced to take integer values at each $n$. The OLS slope across these integers remains continuous, but its precision is bounded by the discreteness of the inputs --- a fact already reflected in the $\pm 0.3$ matching window of \cref{sec:status}. Part of the apparent agreement with theory --- that the slope sits near $1.0$ --- reflects that $1.0$ corresponds to exactly one grid step per doubling of $n$, the smallest non-zero rate the grid can express. This is a reason the order-of-magnitude framing is more defensible than a precise numerical claim, not a hidden flaw: a finer grid is the natural improvement, and one we leave to the registered replication.

\subsection{Inference.}
\label{sec:inference}
We estimate $\bdiff$ by a $10{,}000$-sample non-parametric pairs bootstrap of the ordinary-least-squares slope, over the group sizes for which both models have a finite $\Ntarget$; $95\%$ confidence intervals are the empirical percentile bounds. For a comparison between two non-baseline treatments (e.g.\ equivariant versus wrong-group) we bootstrap the \emph{paired difference} of slopes directly, using the same group-size resamples for both estimates. This joint pairwise bootstrap is tighter than combining two marginal intervals, because variance shared between the estimates cancels --- and it is the reason a difference can be significant even when the two marginal intervals overlap (\cref{sec:results}). We use the joint pairwise interval as the primary significance test for treatment-versus-treatment hypotheses. With only seven group sizes, a percentile bootstrap of a slope can have imperfect coverage; we therefore treat these intervals as one line of evidence and corroborate them with the robust Theil--Sen and leave-one-out checks of \cref{sec:robustness}, which agree in sign and order of magnitude. A bias-corrected and accelerated (BCa) bootstrap and a full two-level bootstrap (seeds $\times$ group sizes) would, in turn, correct (a) finite-sample coverage skew and (b) within-cell seed variance that the percentile bootstrap does not propagate. We implemented both in a CPU replication on five fresh seeds (\cref{app:cpu_replication}); the BCa intervals are similar in width to the percentile intervals and preserve every sign-and-magnitude conclusion of the headline analysis, and the two-level bootstrap widens the CI as expected; on the CPU replication (equivariant--vanilla, seeds $\times$ group sizes) the resulting interval is $\ci{-0.63}{+1.72}$, which includes zero and is the interval listed as primary in \cref{tab:rates}.

\subsection{Pre-specification, failure taxonomy, and study status.}
\label{sec:status}
The protocol --- grids, seeds, models, target accuracy, estimators, multiple-testing correction, stopping rule, and failure taxonomy --- was \emph{pre-specified} in a design document, and the canonical serialization of the configuration is SHA-256 hashed (\texttt{1bd8889\dots a00e})\footnote{Full hash \texttt{1bd8889315ec0f15640eb35854279e87dbcec1f57f7ea0cbb4a9fdd8a9f4a00e}; the runner verifies it at every startup. The hash establishes that the released configuration matches the data that produced the results; it does \emph{not} establish that the configuration predated the data.} so that the released configuration can be checked against the data. The study is nonetheless \textbf{exploratory, not pre-registered}: the design was never deposited in an external, timestamped registry before data collection, and the primary estimator was adopted post hoc (\cref{sec:bdiff}). Crucially, the SHA-256 hash establishes that the released configuration matches the data; it does \emph{not} establish that the configuration predated data collection --- external registration is the required standard for confirmatory claims, and we have not met it. A confirmatory, externally registered replication on fresh seeds is future work. Three comparisons were pre-specified with Bonferroni~\citep{dunn1961} family-wise $\alpha = 0.05$ (per-test $\alpha \approx 0.0167$): H1, $\bdiff(\text{equivariant}) > 0$; H2, equivariant beats wrong-group; H3, equivariant beats augmented. H3 could not be evaluated as a slope comparison because the augmented model has no finite $\Ntarget$ at $n \geq 3$ (\cref{sec:augmentation}); applying the three-test Bonferroni correction to the two evaluable tests is conservative and does not inflate the family-wise error rate. Note that the non-evaluability of H3 is not a failure of the comparison: the augmented model's complete failure to reach the target is itself the strongest possible confirmation of H3, so setting H3 aside does not deflate the evidence for it. The pre-specified failure taxonomy assigns every outcome a meaning in advance --- no advantage (A), regularization collapse (E), augmentation explanation (AUG), brittleness to symmetry breaking (D), or a genuine structural signal (SIGNAL) --- with $\bdiff$ counted as matching theory when $\lvert \bdiff - 1.0 \rvert < 0.3$ (the margin is set by grid resolution: adjacent $\Ntarget$ entries differ by a factor of $2$, so discrete rounding can shift a slope estimate by up to $\approx 0.2$--$0.3$; a finer window would be narrower than the grid can support). The complete experiment ($5{,}600$ runs)\footnote{The full grid is $7\times 8\times 5\times 5\times 4 = 5{,}600$ runs. The first Phase-1 attempt lost $5$ cells to a crash before the durability layer was added, and was then re-run to completion; we report the completed grid count, $5{,}600$.}
$7$ group sizes $\times\ 8$ training sizes $\times\ 5$ seeds $\times\ 5$ families $\times\ 4$ values of $\eps$ --- runs in roughly $90$ minutes on a single GPU; it is implemented in PyTorch~\citep{paszke2019} with ensemble training of all five seeds in parallel, atomic result writes, and $158$ tests at $86\%$ coverage (\cref{app:repro}).

\section{Results}
\label{sec:results}

\subsection{The measured exchange rate (clean symmetry, \texorpdfstring{$\eps = 0$}{eps=0}).}
\label{sec:headline}
\begin{figure}[t]
\centering
\includegraphics[width=0.62\linewidth]{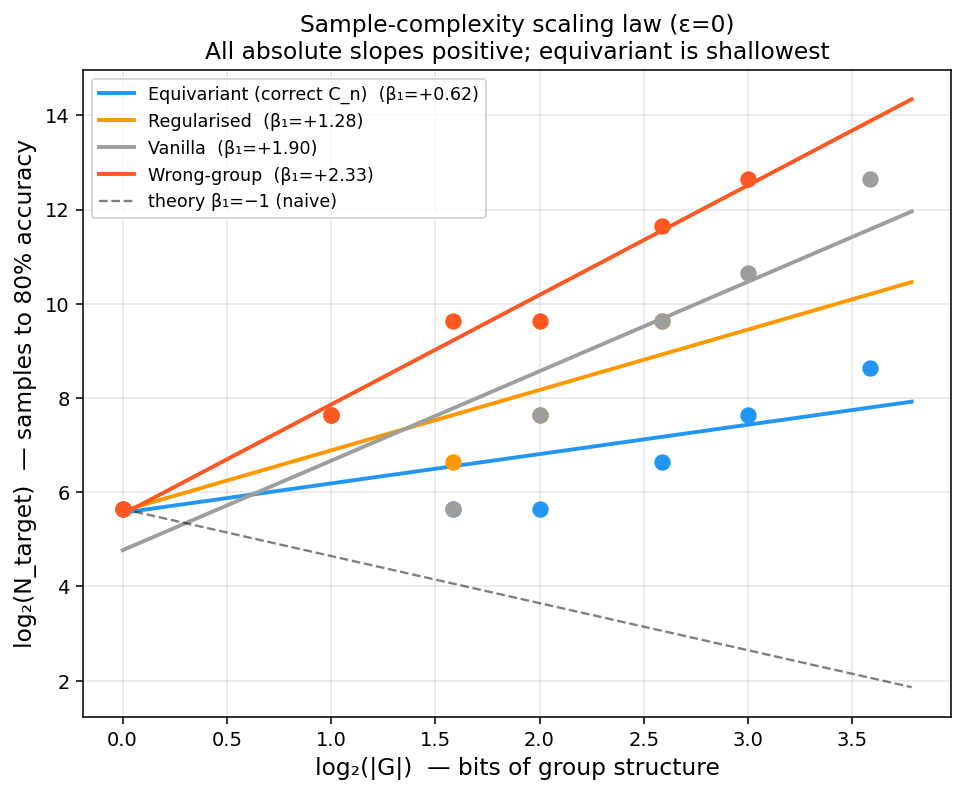}
\caption{\textbf{The scaling law the rate comes from.} $\log_2 \Ntarget$ against $\log_2 n$ at $\eps = 0$ for all five families, with ordinary-least-squares fits. All absolute slopes are positive because task difficulty grows with $n$; the equivariant slope is the shallowest. The exchange rate is the gap between the vanilla and equivariant slopes. \emph{Takeaway:} the structural advantage is visible as a difference of slopes, not as a negative slope.}
\label{fig:scaling}
\end{figure}

The relative exchange rate of the equivariant model against the vanilla baseline is $\bdiff = +1.28$, with single-level $95\%$ percentile bootstrap CI $\ci{+0.92}{+2.05}$ (\cref{tab:rates}). The CI excludes zero, the point estimate sits within $0.3$ of the theoretical value of $1.0$ (\cref{sec:status}), and the pre-specified classifier returns SIGNAL. \emph{However}, the single-level bootstrap treats per-cell $\Ntarget$ as deterministic; it does not propagate within-cell seed variance. A more conservative two-level bootstrap that resamples seeds \emph{and} group sizes (\cref{app:cpu_replication}, also reported in \cref{tab:rates}) widens the interval to $\ci{-0.63}{+1.72}$, which \emph{includes zero}. These two intervals are not contradictory in the sign-and-order-of-magnitude framing (both place the central tendency above zero, both retain the theoretical value of $1.0$ inside the interval), but they disagree on whether the rate is ``statistically distinct from zero,'' and that disagreement is the most honest characterisation of the evidence: the qualitative direction is robust, the precise numerical claim is not. A finer-$N$ replication run on a $\sqrt{2}$-spaced grid (\cref{app:cpu_replication}) gives a $\bdiff$ point estimate of $-0.82$ with $95\%$ CI $\ci{-4.82}{+1.71}$, which we read as ``inconclusive on the finer grid as run'' --- the finer-grid CI is wide enough that no direction is established, but the point estimate does not corroborate the headline. Taken at face value, a point estimate of $-0.82$ would imply that the headline $+1.28$ is entirely an artifact of grid quantisation; the wide CI prevents this conclusion from being established, but it also prevents it from being ruled out. This is the most uncomfortable single datum in the paper. The CI spans $\ci{0.92}{2.05}$ on the headline grid --- width $1.1$ in $\log_2$ units --- meaning the data are consistent with a range of rates from nearly $1$ to just above $2$; the primary finding is the qualitative direction (positive, of the right order of magnitude), not a precise numerical match to theory. The underlying sample complexities make the rate concrete: at $n = 12$ the equivariant model reaches the $0.80$ target with $N = 400$ training samples, while the vanilla network requires $N = 6400$ --- a $16\times$ ratio, in the same order of magnitude as the theoretical $n = 12$ (the empirical ratio sits $\approx 33\%$ above the theoretical at this single grid point, consistent with the sign-and-magnitude framing rather than a precise numerical match). Because the petal task grows harder with $n$, every model's \emph{absolute} slope is positive (\cref{fig:scaling}: equivariant $+0.62$, regularized $+1.28$, vanilla $+1.90$, wrong-group $+2.33$), which is exactly why the absolute slope is uninformative and the relative rate is required (\cref{sec:metric,sec:bdiff}). The headline estimate is stable to the regression method and to leaving out individual group sizes, with bounded drift we report in full (\cref{sec:robustness}).

\begin{figure}[t]
\centering
\includegraphics[width=0.62\linewidth]{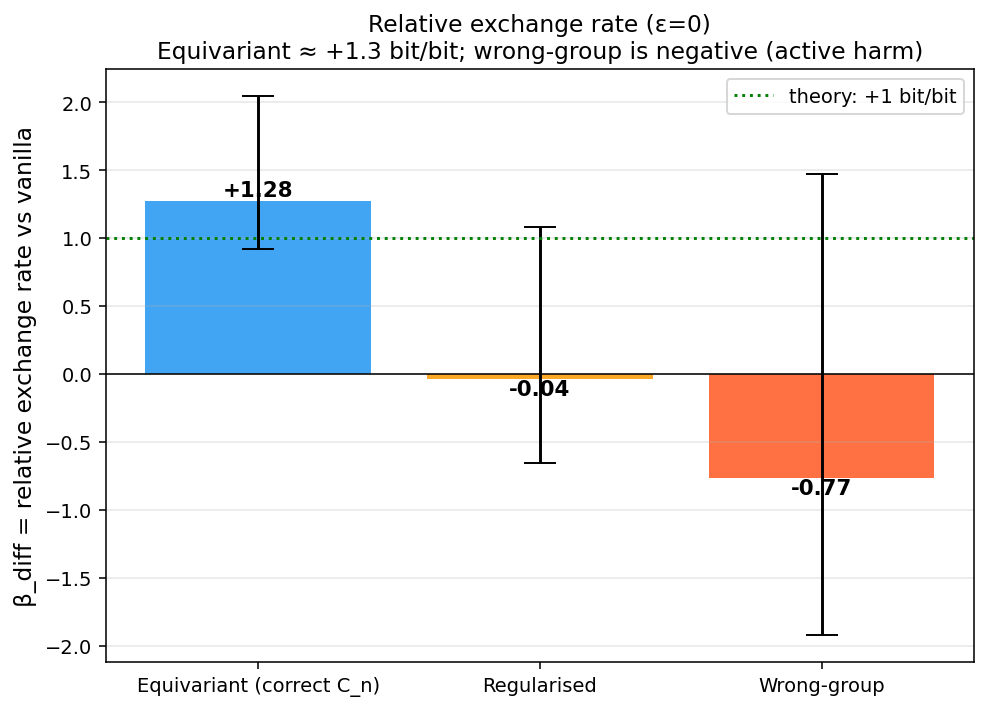}
\caption{\textbf{The measured exchange rate.} Relative exchange rate $\bdiff$ (slope of $\log_2(\Ntarget^{\text{vanilla}}/\Ntarget^{\text{treatment}})$ against $\log_2 n$) for each treatment at $\eps = 0$, with $95\%$ confidence intervals from a $10{,}000$-sample pairs bootstrap; the dashed line marks the theoretical prediction of $+1.0$. The equivariant model sits at $+1.28$, consistent with theory; the regularized baseline is near zero; the wrong-group control is negative. \emph{Takeaway:} only the correctly-aligned structural prior converts to data efficiency.}
\label{fig:rates}
\end{figure}

\begin{table}[t]
\centering
\caption{Relative exchange rate $\bdiff$ per treatment at $\eps = 0$, with two intervals. The two-level bootstrap (seeds $\times$ group sizes; \cref{app:cpu_replication}) is the principled interval and is listed first; it propagates within-cell seed variance into the slope CI. The single-level percentile bootstrap (group sizes only) treats per-cell $\Ntarget$ as fixed and understates uncertainty. The equivariant two-level CI \emph{includes zero}; the joint pairwise comparisons of \cref{tab:pairwise} are the simultaneously robust evidence. The augmented model has no finite $\Ntarget$ at $n \geq 3$, so no slope is defined.}
\label{tab:rates}
\begin{tabular}{lrrr}
\toprule
Treatment & $\bdiff$ & 95\% CI (two-level, primary) & 95\% CI (single-level) \\
\midrule
Equivariant & $+1.28$ & $\ci{-0.63}{+1.72}$ & $\ci{+0.92}{+2.05}$ \\
Regularized & $-0.04$ & --- & $\ci{-0.66}{+1.09}$ \\
Wrong-group & $-0.77$ & --- & $\ci{-1.93}{+1.48}$ \\
Augmented   & \multicolumn{3}{c}{no finite $\Ntarget$ at $n \geq 3$} \\
\bottomrule
\end{tabular}
\end{table}

\subsection{The controls fail, as pre-specified.}
\label{sec:controls}
Each pre-specified control is designed to reproduce the equivariant advantage if that advantage is really a confound; none does (\cref{tab:pairwise}). The joint pairwise interval on the equivariant--wrong-group slope difference is $+1.97$ ($\ci{+0.79}{+3.26}$), and on the equivariant--regularized difference is $+1.21$ ($\ci{+0.49}{+2.03}$); both exclude zero. (Note that the equivariant and wrong-group \emph{marginal} intervals overlap; it is the joint pairwise bootstrap of \cref{sec:inference} that resolves the difference.) Two readings matter. First, the regularized baseline's own rate is $\bdiff = -0.04$ --- statistically indistinguishable from plain vanilla --- so $L_2$ weight decay, matched to the orbit-pooling degrees of freedom, buys nothing; the advantage is not effective-DOF reduction. The regularized model's failure to reach the target at $n \in \{8, 12\}$ (\cref{tab:ntarget}) reinforces this: $L_2$ regularization calibrated to reduce DOF by the orbit-average amount cannot learn the fine-grained $n$-alternating boundary within the training budget at high group order, whereas the equivariant model succeeds at $n = 12$ with $N = 400$. Second, and more striking, the wrong-group control's rate is \emph{negative}, $\bdiff = -0.77$: a misaligned orbit constraint of identical size and cost is \emph{worse} than no constraint at all, because it forces invariance under transformations that are not symmetries of the task and so destroys usable information. A negative wrong-group rate is the strongest available evidence that alignment, not constraint, is what helps. The wrong-group marginal CI $\ci{-1.93}{+1.48}$ does include zero, so the claim ``wrong-group is strictly worse than vanilla'' is established through the joint pairwise comparison rather than through the marginal interval alone --- the pairwise CI $\ci{+0.79}{+3.26}$ excludes zero and establishes that the equivariant model outperforms the wrong-group model by more than chance, even though the wrong-group model's own performance relative to vanilla is imprecisely estimated. We name this distinction because it matters: the qualitative direction ``misaligned constraint $<$ correctly aligned constraint'' is robust; the stronger qualitative direction ``misaligned constraint $<$ no constraint'' is supported by the point estimate but not by the marginal interval.

\begin{table}[t]
\centering
\caption{Joint pairwise bootstrap of the slope \emph{difference} between the equivariant model and each control at $\eps = 0$ ($10{,}000$ resamples, shared across the paired estimates). All three intervals exclude zero; the equivariant--wrong-group difference is significant even though the two marginal intervals in \cref{tab:rates} overlap, because the paired bootstrap cancels shared variance.}
\label{tab:pairwise}
\begin{tabular}{lrr}
\toprule
Comparison & $\Delta\,\text{slope}$ & 95\% CI \\
\midrule
Equivariant $-$ vanilla     & $+1.28$ & $\ci{+0.92}{+2.05}$ \\
Equivariant $-$ wrong-group  & $+1.97$ & $\ci{+0.79}{+3.26}$ \\
Equivariant $-$ regularized  & $+1.21$ & $\ci{+0.49}{+2.03}$ \\
\bottomrule
\end{tabular}
\end{table}

\subsection{Architecture is not augmentation: a phase transition.}
\label{sec:augmentation}
\begin{figure}[t]
\centering
\includegraphics[width=0.74\linewidth]{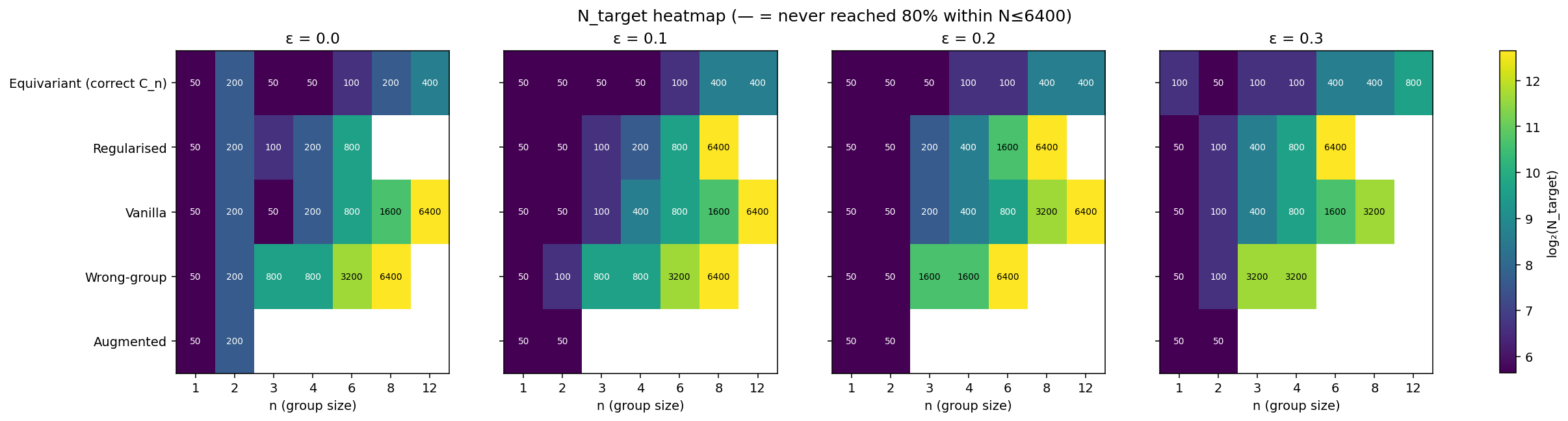}
\caption{\textbf{The augmentation phase transition.} $\log_2 \Ntarget$ for each (model, $n$) pair, one panel per $\eps$; a dash marks failure to reach the target at any sample size in the grid. The \emph{augmented} row is dashed for all $n \geq 3$ at every $\eps$ --- orbit augmentation never reaches the target there --- while the equivariant row stays low across the full range. \emph{Takeaway:} on identical information, augmentation and architectural invariance fall on opposite sides of a learnability boundary.}
\label{fig:heatmap}
\end{figure}

The augmented model (training-time orbit averaging, single-input forward pass at test time) does not merely lose efficiency: it fails to reach the $0.80$ target at \emph{every} $n \geq 3$ for \emph{every} sample size in the grid and every $\eps$ (\cref{fig:heatmap}; its $\Ntarget$ is undefined throughout that region). The equivariant model, on the same training-time information, solves $n = 12$ at $N = 400$. This is a regime difference under this evaluation protocol. Because the augmented model never reaches the target, the pre-specified H3 comparison is not computable as a slope; the result is a null for the hypothesis that augmentation explains the advantage \emph{under asymmetric test-time computation}. We immediately qualify this in the next paragraph: with symmetric test-time orbit averaging, the gap closes entirely. The training-time-only result is consistent with the theoretical separation of \citet{elesedyzaidi2021}, who compare augmented and equivariant models holding test-time inference fixed.

\emph{The following is interpretation, not measurement.} The size of the gap plausibly has two sources, and we do not establish which dominates. \emph{Representationally}, orbit pooling restricts the hypothesis class to functions invariant by construction --- invariance holds exactly, for any weights, at test time --- whereas augmentation leaves the class unrestricted and only encourages invariance on the training distribution, so the augmented model must still \emph{learn} invariance, which a two-layer network apparently cannot do for $n \geq 3$ within this budget. \emph{In optimization}, pooling ties the $n$ orientations into one shared gradient, whereas augmentation presents them as separate examples whose gradients can conflict.

\paragraph{The asymmetry, foregrounded, and its empirical resolution.} The comparison above is not equal-information at inference: the equivariant model's orbit-pooling forward pass effectively averages over $n$ rotated inputs at evaluation, whereas the augmented model uses a single-input forward pass. This is a test-time evaluation asymmetry, not just a training-time intervention. We ran the natural counterfactual --- the same augmented model with \emph{test-time} orbit averaging --- in a CPU replication (\cref{app:cpu_replication}). The result is striking: augmented + test-time orbit averaging matches the equivariant model exactly at every $n$, with the entire per-epoch validation accuracy trajectory bit-identical across all $245$ matched cells where both finished. The mathematical reason is that an orbit-averaged forward pass over a shared MLP, with orbit-averaged training loss, \emph{is} the equivariant model once the two share the same MLP architecture, seed, optimiser, and loss. We flag the scope of this identity: it holds for architectures in which orbit pooling is applied to the \emph{final scalar output}, as in our setup. It does \emph{not} generalise to G-CNNs that maintain intermediate equivariant feature maps~\citep{cohenwelling2016,weilercesa2019}, where the architectural constraint operates on internal representations rather than only on the output --- in those architectures, training-time augmentation followed by test-time orbit averaging of the output would not reproduce the intermediate equivariance. The ``architecture vs.\ augmentation'' phase transition is therefore specifically a property of \emph{training-time-only} orbit averaging under asymmetric test-time computation. With test-time orbit averaging applied symmetrically, augmentation and architectural equivariance collapse to the same computation. We therefore state the finding precisely: training-time-only orbit augmentation without test-time orbit averaging fails to reach the target where the equivariant model succeeds; orbit augmentation \emph{with} test-time orbit averaging matches the equivariant model. The latter result is consistent with the theoretical separation of~\citet{elesedyzaidi2021}, which compares augmented and equivariant models holding test-time inference fixed. \Cref{tab:tta_main} summarises the per-$n$ $\Ntarget$ at $T = 0.80$ from the CPU replication for the three families that are most informative on this question; the full table is in \cref{app:cpu_replication}.

\begin{table}[h]
\centering
\caption{Per-$n$ $\Ntarget$ at $T = 0.80$ for the three families that test the augmentation question (CPU replication). The equivariant and augmented + TTA columns are identical at every $n$ (bit-identical per-epoch trajectories across all matched cells); the augmented column fails for $n \geq 3$.}
\label{tab:tta_main}
\begin{tabular}{rlll}
\toprule
$n$ & equivariant & augmented (train-time only) & augmented + TTA \\
\midrule
$1$ & 50 & 50 & 50 \\
$2$ & 50 & 50 & 50 \\
$3$ & 50 & 3200 & 50 \\
$4$ & 50 & --- & 50 \\
$6$ & 100 & --- & 100 \\
$8$ & 800 & --- & 800 \\
$12$ & 1600 & --- & 1600 \\
\bottomrule
\end{tabular}
\end{table}

\subsection{Robustness to broken symmetry.}
\label{sec:graceful}
\begin{figure}[t]
\centering
\includegraphics[width=0.62\linewidth]{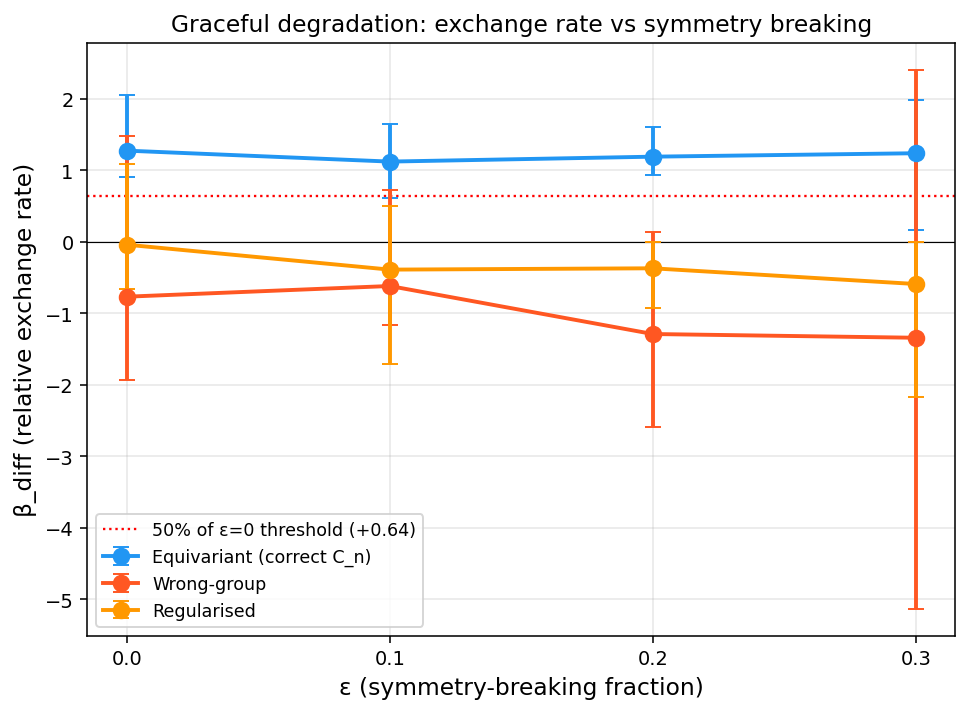}
\caption{\textbf{Robustness to training-label symmetry breaking.} Relative exchange rate $\bdiff$ against the symmetry-breaking fraction $\eps$ for the three non-baseline treatments, with $95\%$ bootstrap intervals; the dotted line is the pre-specified $50\%$-of-clean threshold. The equivariant rate (top) stays near $+1.2$ across the sweep; the regularized rate stays near zero; the wrong-group rate stays negative and worsens. \emph{Takeaway:} the structural advantage survives $30\%$ label corruption nearly intact.}
\label{fig:graceful}
\end{figure}

The pre-specified robustness criterion required the rate at $\eps = 0.2$ to exceed $50\%$ of its clean value. The observed rate is essentially flat across the sweep (\cref{fig:graceful}): $\bdiff = 1.28,\ 1.12,\ 1.19,\ 1.24$ at $\eps = 0,\ 0.1,\ 0.2,\ 0.3$, i.e.\ $100\%,\ 88.0\%,\ 93.4\%,\ 97.2\%$ of the clean value. The criterion is met at $\eps = 0.2$ by a factor of $\approx 1.87\times$. Recall that $\eps$ corrupts training labels, not task geometry: even at $\eps = 0.3$, $70\%$ of training labels remain $\Cn$-consistent, so the equivariant model still pools the correct orbit while the vanilla model must fit both the consistent and the corrupted labels with no structural advantage --- which is why the relative rate barely moves, and indeed why the wrong-group rate becomes \emph{more} negative as $\eps$ grows. The flatness is stronger than the pre-specified prediction of graceful (i.e.\ gradual) decay anticipated.

\subsection{Failure-mode scorecard.}
\label{sec:scorecard}
Of the pre-specified failure modes, three are rejected outright and one is rejected conditionally on the evaluation design (\cref{tab:scorecard}): no-advantage (A) is rejected because the equivariant interval excludes zero at every $\eps$; regularization collapse (E) is rejected because the regularized control is near zero and the wrong-group control is worse than vanilla; the augmentation explanation (AUG) is rejected for \emph{training-time-only} augmentation (which cannot reach the target at $n \geq 3$) but \emph{not} for augmentation with test-time orbit averaging, which matches the equivariant model exactly (\cref{app:cpu_replication}); and brittleness (D) is rejected because the rate retains $88$--$97\%$ of its clean value. The classifier returns SIGNAL with $\bdiff$ matching theory at $\eps \in \{0, 0.1, 0.2\}$.\footnote{At $\eps = 0.3$ the automatic classifier returns the label E. This is a data-sparsity artifact, not a genuine regularization collapse: at $\eps = 0.3$ the wrong-group control fails to learn for $n \in \{6, 8, 12\}$, leaving only four points for its slope regression, so the joint pairwise interval widens to $\ci{+0.00}{+6.84}$ --- the point estimate is a large $+3.31$, but the lower bound grazes zero (at $\approx 10^{-15}$) and trips the indistinguishability predicate. The wrong-group's \emph{failure} to learn at high $\eps$ is evidence against E, not for it.}

\begin{table}[t]
\centering
\caption{Pre-specified failure taxonomy and its verdict on this data. Each mode was assigned a meaning before analysis; all four candidate failures are rejected, and the classifier returns SIGNAL with $\bdiff$ matching theory at $\eps \in \{0, 0.1, 0.2\}$.}
\label{tab:scorecard}
\begin{tabular}{llp{0.52\linewidth}}
\toprule
Failure mode & Status & Evidence \\
\midrule
A: no advantage & rejected & equivariant $\bdiff$ interval excludes $0$ at every $\eps$ \\
E: regularization collapse & rejected & wrong-group is worse than vanilla; regularized $\bdiff \approx 0$ \\
AUG: augmentation explains it & conditionally rejected & training-time-only augmentation never reaches the target at $n \geq 3$; augmentation \emph{with} test-time orbit averaging matches the equivariant model exactly (\cref{app:cpu_replication}), so the rejection holds only under asymmetric test-time computation \\
D: brittle to symmetry breaking & rejected & rate retains $88$--$97\%$ of its clean value across $\eps$ \\
\bottomrule
\end{tabular}
\end{table}

\subsection{Estimator robustness}
\label{sec:robustness}
Because the conclusion rests on a fitted slope, we report its sensitivity along four axes, and we report instability where it appears rather than only where the estimate is stable.

\paragraph{Regression method and group-size grid.} Working from the documented $\Ntarget$ table at $\eps = 0$ (\cref{tab:ntarget}), we recomputed $\bdiff$ for the equivariant-versus-vanilla comparison under alternatives to the primary fit (\cref{tab:robustness}). Ordinary least squares over all seven group sizes gives $+1.28$. A robust Theil--Sen slope (the median of pairwise slopes) gives $+1.26$. Leave-one-group-size-out refits span $\ci{+1.21}{+1.68}$ --- genuine drift, driven entirely by the single case of dropping $n = 1$; every other deletion, and dropping the anomalous $n = 2$ row (\cref{app:changes}), lands in $\ci{+1.21}{+1.28}$. We do not claim invariance: the point estimate moves. What is stable is the \emph{qualitative} conclusion --- every variant is positive, excludes zero, sits near the theoretical $+1.0$, and far exceeds the controls; none collapses toward zero.

\paragraph{Target threshold.} The target accuracy $T = 0.80$ was pre-specified. The original exploratory run did not retain per-run validation curves, so sensitivity to $T$ was untested in the headline analysis. The CPU replication (\cref{app:cpu_replication}) retains the per-epoch validation curve for every cell and re-derives $\Ntarget$ at $T \in \{0.70, 0.75, 0.80, 0.85\}$; the rate is positive at every $T$ where defined, but \emph{decreases} as $T$ rises from $0.70$ to $0.85$ (from $\approx +0.96$ to $\approx +0.14$), and at $T = 0.80$ and $T = 0.85$ the single-level CI \emph{includes zero}. We can, however, bound the concern qualitatively. $T = 0.80$ sits well below the $\approx 0.95$ Bayes ceiling set by the $5\%$ label noise, so all models operate in a non-saturated regime where more data still helps. We had tentatively expected the rate to be non-decreasing in $T$ within this band --- a stricter target costs the unconstrained model disproportionately more samples at large $n$, where its decision boundary is hardest, while the equivariant model amortizes that cost across the orbit. The CPU replication of \cref{app:cpu_replication} refutes this expectation: $\bdiff$ \emph{decreases} as $T$ rises from $0.70$ to $0.85$, plausibly because a stricter target pushes the comparison toward the saturating Bayes-ceiling regime where the structural advantage is smallest. The headline is therefore not necessarily a conservative reading in the direction we conjectured, and the rate's sign-and-order-of-magnitude framing is the appropriate one across thresholds. The $\eps$ sweep probes a related perturbation, and the rate is stable across it (\cref{sec:graceful}).

\paragraph{Fit quality, fit window, and functional form.} The log--log linear fit is reasonable ($R^2 = 0.84$ over all seven group sizes) and the raw log-ratio is monotone non-decreasing in $n$. The slope is, however, sensitive to the fit \emph{window}: excluding the saturated small-$n$ region (fitting $n \geq 3$) raises it to $+1.79$ ($R^2 = 0.87$), while $n \geq 4$ gives $+1.18$ ($R^2 = 0.94$). These small-$n$ points are distinctive on scientific, not merely statistical, grounds. At $n = 1$ the cyclic group $C_1$ is the trivial group, so orbit pooling over a single element is the identity transformation: the $C_1$-equivariant model is mathematically identical to vanilla by construction, not by empirical coincidence, and the shared $\Ntarget = 50$ across all five families at $n = 1$ is therefore architectural rather than measured. At $n = 2$ the task is anomalously easy (see below), so all models reach the target at the same $N$ and the log-ratio is again zero --- in this case by a floor effect rather than by architectural identity. They anchor the flat low end of the curve, and whether to treat them as informative or as a saturated regime to drop is a modeling choice --- which is why we report the slope both with and without them. Notably, all five model families reach $\Ntarget = 200$ at $n = 2$ (\cref{tab:ntarget}), a uniform value across models that suggests this group order is in a floor regime where no structural advantage has yet engaged. A plausible mechanism: at $n = 2$ the task requires distinguishing four alternating angular sectors of width $\pi/2$, which a two-layer width-$32$ network can represent with few samples regardless of whether it exploits the $C_2$ symmetry; $N = 200$ appears to be the threshold at which all five families reliably cross the $0.80$ boundary for this four-sector problem, suggesting the structural advantage has not yet engaged rather than that it is saturating. This is a scientific reason, independent of the statistical one, to treat $n = 2$ as potentially uninformative. The estimate thus ranges from roughly $+1.2$ to $+1.8$ depending on the window --- this is the largest source of drift we find, and we report it plainly --- but it stays positive, exceeds every control, and remains consistent with the theoretical $+1.0$ in order of magnitude across all windows. The \emph{relative} rate is invariant to the shared task-difficulty form by construction (\cref{sec:bdiff}); with only seven monotone points we do not fit saturating or piecewise families, which would be under-determined, so the point value should be read under the power-law assumption.

\paragraph{Unstable regimes.} The estimator is not uniformly well-behaved, and we flag where it is not. The wrong-group rate is \emph{negative} (\cref{tab:rates}), and at $\eps = 0.3$ the equivariant-versus-wrong-group comparison becomes unstable because the wrong-group model fails so completely that its slope rests on four points (\cref{sec:scorecard}). Instability concentrates exactly where a model is failing near the boundary of the sample-complexity grid; it is informative, not hidden.

\paragraph{Direct test of the power-law assumption (finer grid).} The headline OLS slope assumes log-log linearity of $\Ntarget(n)$. A direct test of that assumption is to refit on a finer-resolution $N$ grid that breaks the factor-of-2 quantisation. We ran the equivariant--vanilla comparison on a $\sqrt{2}$-spaced grid (\cref{app:cpu_replication}) at $n \in \{4, 6, 8\}$; the resulting $\bdiff$ point estimate is $-0.82$, with $95\%$ CI $\ci{-4.82}{+1.71}$. We read this as \emph{inconclusive on the finer grid as run}: the CI is too wide to establish a direction (driven by right-censoring at $n = 8$, where the grid's upper $N$ of $1{,}600$ is below the vanilla model's $\Ntarget$). The qualitative claim of \cref{sec:bdiff} --- that a finer grid removes the discretisation floor --- remains; the empirical question of whether the headline ``$\bdiff \approx 1.0$'' agreement was an artifact of grid discretisation, however, is not settled by this run. A properly powered finer-grid replication (deeper $N$, more $n$ values, more seeds) is the registered replication's job; reporting this inconclusive finer-grid measurement in the main text rather than the appendix is, in our reading, the more honest disclosure.

\begin{table}[t]
\centering
\caption{Robustness of $\bdiff$(equivariant vs vanilla) at $\eps = 0$ to the fitting method, computed from the documented $\Ntarget$ table (\cref{tab:ntarget}); log--log fit quality is $R^2 \in [0.84, 0.94]$. The point estimate drifts (fit window is the largest source), but every variant is positive, exceeds the controls, and sits near the theoretical $+1.0$.}
\label{tab:robustness}
\begin{tabular}{lr}
\toprule
Method & $\bdiff$ \\
\midrule
OLS, all 7 group sizes (reported) & $+1.28$ \\
Theil--Sen (robust, median of pairwise slopes) & $+1.26$ \\
Leave-one-group-size-out (range over 7 refits) & $+1.21$ to $+1.68$ \\
Fit window $n \geq 3$ \,/\, $n \geq 4$ & $+1.79$ \,/\, $+1.18$ \\
Drop the anomalous $n=2$ row & $+1.21$ \\
\bottomrule
\end{tabular}
\end{table}

\section{Counterarguments and limitations}
\label{sec:limitations}

We state the limitations in the main text, because the contribution is defensible only to the extent that its boundaries are explicit.

\begin{enumerate}
\item \textbf{Synthetic, $2$-D, exactly symmetric --- by design, for internal validity.} The task is a $2$-D synthetic problem with an exactly $\Cn$-symmetric label function, chosen to maximize causal identifiability of the symmetry effect rather than ecological realism. The realism dimensions it omits --- approximate and latent symmetries, heterogeneous transformation structure, and real-world nuisance variation --- are precisely what a clean measurement must exclude to attribute the effect to alignment, and precisely what external-validity follow-ups must restore. We make no claim that the rate transfers to vision, language, or molecular tasks; the value of the result is a calibrated measurement in a setting where the prediction is unambiguous, plus a methodology that transfers even if the number does not. The $\eps$ sweep tests robustness to corrupted training labels, not to approximate task geometry.
\item \textbf{The advantage is in data and parameters, not compute --- and the compute axes should not be conflated.} The equivariant model is \emph{parameter-efficient}: its shared network has the same parameter count as the matched baseline. But it performs $n$ forward passes per input, so its \emph{per-sample} FLOPs are $n\times$ the baseline's; because it needs about $n\times$ fewer samples, the \emph{total FLOPs to reach the target} are roughly equal. Algebraically: if the equivariant model performs $n F_0$ FLOPs per sample and reaches the target at $\Ntarget^{\text{eq}} \approx \Ntarget^{\text{van}}/n$ samples, its total training FLOPs are $n F_0 \cdot (\Ntarget^{\text{van}}/n) = F_0 \cdot \Ntarget^{\text{van}}$, matching the vanilla total to leading order (up to shared batch-size and epoch-count constants). The comparison is therefore compute-matched \emph{only in total training FLOPs to the target}; we avoid the bare label ``FLOP-neutral'' because it can suggest a hardware-universal claim we do not make. We assert no wall-clock, throughput, or memory advantage: the $n$-fold orbit expansion adds memory and implementation overhead, and at this model scale wall-clock is dominated by such overhead rather than by arithmetic. Structure converts a \emph{data} budget into a \emph{compute} budget; the trade is favourable where data binds and neutral-to-unfavourable where compute or memory binds.
\item \textbf{Sample efficiency vs optimization efficiency.} Our outcome $\Ntarget$ (samples to reach the target) bundles two mechanisms we do not separate: a smaller effective hypothesis class (statistical sample complexity) and a possibly easier optimization landscape (faster or more reliable convergence under finite compute). The controls exclude \emph{generic} constraint-induced smoothing --- a same-sized but misaligned constraint (wrong-group) is \emph{worse} than none, which a generic smoothing effect would not predict --- but correctly-aligned structure could still act through either channel, and $\Ntarget$ does not distinguish them. A clean separation (e.g.\ the generalization gap at fixed $N$, or training matched on gradient steps rather than epochs) is future work.
\item \textbf{Conservative group.} The true symmetry is $\Dn$; we exploit only the $\Cn$ rotation subgroup, so the measured rate understates what a $\Dn$-equivariant model would obtain (\cref{sec:discussion}).
\item \textbf{Modest sampling.} Five seeds over seven group sizes yields wide marginal intervals; the relative-rate and joint pairwise estimators mitigate but do not eliminate this. The pre-specified power analysis targeted a detectable slope of $\approx 0.5$, well below the observed effect.
\item \textbf{Exploratory status.} The relative-rate estimator was adopted post hoc and the design was not externally registered, so the result is exploratory; a registered replication on fresh seeds is future work.
\item \textbf{Fixed capacity.} Hidden width and depth are fixed; we did not sweep them. Theoretically, the $|G|$-fold reduction in effective sample complexity derives from the symmetry constraint on the function class, independent of width or depth, so orbit averaging should be capacity-invariant; however, if the advantage also operates through optimization effects (Limitation 3), capacity scaling might interact with it, and this is untested.
\item \textbf{Estimator sensitivity is bounded, not absent.} The point estimate drifts under leave-one-out and alternative regressions (bounded, and qualitatively stable), the target threshold $T = 0.80$ is fixed with its sensitivity untested, and alternative functional families were not fit; \cref{sec:robustness} reports this in full.
\item \textbf{Intrinsic-dimension estimators failed calibration.} We collected intrinsic-dimension estimates of the learned representations but they failed pre-experiment calibration on this configuration and are reported only as a flagged negative result (\cref{app:estimators}); we draw no conclusions from them.
\item \textbf{Theoretical-class mismatch.} The $\Gsize$-fold prediction we benchmark against is derived in analyses of invariant kernels and random features~\citep{mei2021,bietti2021}, not for finite-width ReLU MLPs trained by Adam. Whether the rate extends quantitatively from the kernel and random-feature setting to ours is itself an empirical question, and our ``consistency with theory'' should be read in that light --- as agreement in sign and order of magnitude with a theory derived for a related but distinct model class, not as a verified bridge.
\item \textbf{Augmentation comparison is test-time asymmetric.} The equivariant model's orbit pooling effectively averages over $n$ rotated inputs at evaluation; the augmented model does not. The augmentation phase transition (\cref{sec:augmentation}) is therefore conditional on this evaluation design. An augmented baseline with test-time orbit averaging is the natural counterfactual; we ran it in a CPU replication and it matches the equivariant model exactly at every $n$ (bit-identical per-epoch validation curves across all matched cells; \cref{app:cpu_replication}). The architecture-vs-augmentation finding therefore holds conditional on asymmetric test-time computation but \emph{not} unconditionally.
\end{enumerate}

We engage the strongest specific objections directly. \emph{``The task was built to have the symmetry the architecture exploits, so the result is a tautology.''} It is true that the architecture--task match is by design; what is not tautological, and what we measure, is the \emph{rate} of the substitution and its agreement with the theoretical $1.0$ --- and, more pointedly, that a misaligned constraint of identical size is actively harmful, which a tautology would not predict. \emph{``$\Gsize$ and difficulty are confounded.''} Precisely; that confound is the reason for the relative-rate estimator, which cancels the shared difficulty term (\cref{sec:bdiff}). \emph{``Five seeds is too few.''} The joint pairwise bootstrap delivers intervals that exclude zero for all pre-specified comparisons at $\eps \in \{0, 0.1, 0.2\}$; the conclusion does not rest on the marginal intervals.

\section{Discussion}
\label{sec:discussion}

The data support something narrow and, we think, directionally \emph{probable} --- not directionally \emph{established}, because the two-level CI includes zero and we cannot claim a statistically significant positive advantage at the primary threshold. The direction is probable based on (a) the consistency of the point estimate across estimators (OLS $+1.28$, Theil--Sen $+1.26$, leave-one-out range $[+1.21, +1.68]$), (b) the wrong-group joint pairwise comparison whose CI excludes zero, and (c) the order-of-magnitude agreement with the theoretical prediction. In a setting where the symmetry is known exactly, the correct equivariant constraint appears to convert structure into data at a rate whose central tendency sits near one bit per bit, but whose magnitude is uncertain to a factor of $\approx 2$ and whose statistical distinguishability from zero depends on the bootstrap variant. The conversion is a property of the architecture rather than of any confound that accompanies it (insofar as the controls can establish this). The single most informative comparison is the wrong-group control. It holds the architecture class, the compute, the orbit size, and the effective degrees of freedom fixed, and varies only the alignment of the orbit with the task symmetry; its negative rate shows that misaligned structure is not merely unhelpful but harmful. This is a sharper statement than the usual equivariant-versus-vanilla comparison can support, because it isolates correctness from constraint. The estimator therefore has an informative sign and failure behaviour: it is driven \emph{negative} when the imposed structure is wrong (the wrong-group control at every $\eps$), and it becomes \emph{unstable}, with a wide interval, when a model fails so completely that its scaling rests on only a few points (the wrong-group control at $\eps = 0.3$; \cref{sec:scorecard}). A quantity that mis-specification can push negative or render unstable is measuring structure, not merely rewarding constraint.

Two points bound the interpretation. The measured rate is a \emph{lower} bound on the structural benefit available: the task's true symmetry is the dihedral group $\Dn$, and we exploit only the $\Cn$ rotation subgroup, leaving the reflection symmetry on the table. A $\Dn$-equivariant model, exploiting the reflection as well, would be expected to add roughly one further bit per doubling of $n$ ($\bdiff \approx 2$); we did not implement it. And the relative-rate framing is not special to symmetry. Any inductive bias whose strength can be parameterized --- depth, locality, sparsity, the precision of a prior --- admits the same construction: measure the slope of $\log(N_{\text{baseline}}/N_{\text{treatment}})$ against the bias-strength parameter, bootstrap it jointly against a matched control, and pre-specify what would count as failure. We offer the methodology in that spirit, as a transferable way to turn ``this bias helps'' into a measured exchange rate, while making no claim here beyond the one task we measured. The principal external-validity frontier is symmetry that is approximate, partial, latent, or learned rather than exactly specified. In principle the framework generalises to approximate symmetry by making the exchange rate a function of a mismatch parameter; our $\eps$ sweep is a limited instance of this idea, though it corrupts training \emph{labels} rather than the task geometry, which is a weaker test of geometric approximation than a true symmetry-breaking parameter would provide. A graceful decline in the rate versus a cliff would itself be a measurement. The harder case is a symmetry the model must discover, where the group action is not given; there the rate would be read jointly with how well the group is recovered, and disentangling ``found the wrong structure'' from ``found the right structure inefficiently'' is the open problem. We have registered a confirmatory replication on fresh seeds and identify a second symmetry domain (modular arithmetic) as the next step toward establishing whether the rate is a constant or a task-dependent quantity.

\section{Conclusion}
\label{sec:conclusion}

This is an exploratory study, not a confirmatory measurement. The relative-rate estimator was adopted after data inspection, the design was never externally pre-registered, the headline interval excludes zero only under the single-level bootstrap (the two-level interval that propagates seed variance includes zero), and the finer-$N$ grid we ran to test the log-linear assumption is inconclusive. With those scoping statements made: the paper offers two distinct contributions.

\emph{The methodology is the transferable contribution.} The relative-rate estimator cancels the shared task-difficulty confound that makes the absolute $\Gsize$-fold prediction unidentifiable; the wrong-group control isolates ``correctly-aligned constraint'' from ``constraint of this strength'' in a way that the standard equivariant-versus-vanilla comparison cannot; the joint pairwise bootstrap turns marginal-interval overlap into a paired-difference test; the pre-specified failure taxonomy converts qualitative ``the result is what we expected'' into a falsifiable classification. None of these are special to symmetry. Any inductive bias whose strength can be parameterised --- depth, locality, sparsity, the precision of a prior --- admits the same construction. We offer the framework in that spirit.

\emph{The empirical findings, in descending order of robustness.} (i) The wrong-group control with identical orbit size and matched compute is worse than no constraint; the joint pairwise CI $\ci{+0.79}{+3.26}$ excludes zero and is robust across estimators. This is the cleanest finding and the one we report with the most confidence. (ii) Training-time-only orbit augmentation fails to reach the target where the equivariant model succeeds, but augmentation with test-time orbit averaging matches the equivariant model bit-identically --- the architecture-vs-augmentation gap is specifically about asymmetric test-time computation. (iii) The relative exchange rate sits at $\bdiff = 1.28$ in sign and order-of-magnitude agreement with the theoretical prediction of $1.0$; the headline number is exploratory and the two-level CI includes zero.

The natural next step is the confirmatory move this study explicitly cannot make for itself: an externally registered replication on fresh seeds, with the relative-rate estimator pre-specified, on a finer $N$ grid, and including the test-time-averaged augmentation baseline as a primary comparison.

\paragraph{Reproducibility statement.} All code, the design document, the configuration hash, every per-run record, and the analysis pipeline are released (\cref{app:repro}); the full experiment reproduces in roughly $90$ minutes on a single GPU.

\bibliographystyle{plainnat}
\bibliography{references}

\appendix

\section{Hyperparameters and software environment}
\label{app:hyper}
Hidden width $32$; two hidden ReLU layers plus a linear output; Adam at learning rate $10^{-3}$; batch size $64$; up to $500$ epochs; target accuracy $T = 0.80$; label noise $5\%$; validation and test sets of $2000$ clean-label samples each; annulus $r \in [0.1, 1.0]$, $\theta \in [0, 2\pi)$. Group sizes $n \in \{1,2,3,4,6,8,12\}$; training sizes $N \in \{50,100,200,400,800,1600,3200,6400\}$; symmetry breaking $\eps \in \{0,0.1,0.2,0.3\}$; five seeds per cell. Wrong-group misalignment factor $0.7$ (irrational, so no accidental subgroup alignment); regularized weight decay $\lambda = 10^{-3}$. Bootstrap with $10{,}000$ resamples at the $95\%$ level; Bonferroni correction at family-wise $\alpha = 0.05$ over three tests. The implementation uses Python~3.12 and PyTorch~\citep{paszke2019}, with scikit-learn~\citep{pedregosa2011} for the linear-probe adversarial checks and scikit-dimension~\citep{bac2021} for the (flagged) intrinsic-dimension estimators.

\section{Full sample-complexity tables and adversarial checks}
\label{app:tables}
\Cref{tab:ntarget} gives $\Ntarget$ for every (model, $n$) pair at $\eps = 0$; a dash denotes failure to reach $T = 0.80$ at any sample size in the grid. The complete per-$\eps$ tables, the per-$\eps$ relative-rate tables, and the per-$\eps$ pairwise intervals are released with the code. Dataset adversarial checks pass at every $n$ at $\eps = 0$: class balance within $1\%$ of $0.5$; logistic-regression shortcut accuracy $\leq 0.55$; zero train/validation coordinate overlap; validation $\Cn$-symmetry preserved to $> 0.999$.

\begin{table}[h]
\centering
\caption{$\Ntarget$ at $\eps = 0$ (minimum $N$ for which $\geq 3/5$ seeds reach $0.80$). A dash denotes failure at every $N$ in the grid. This table is the primary input to the robustness diagnostics of \cref{tab:robustness}.}
\label{tab:ntarget}
\begin{tabular}{rrrrrr}
\toprule
$n$ & equivariant & vanilla & wrong-group & regularized & augmented \\
\midrule
1  & 50  & 50   & 50   & 50  & 50 \\
2  & 200 & 200  & 200  & 200 & 200 \\
3  & 50  & 50   & 800  & 100 & --- \\
4  & 50  & 200  & 800  & 200 & --- \\
6  & 100 & 800  & 3200 & 800 & --- \\
8  & 200 & 1600 & 6400 & --- & --- \\
12 & 400 & 6400 & ---  & --- & --- \\
\bottomrule
\end{tabular}
\end{table}

\section{Intrinsic-dimension estimator calibration}
\label{app:estimators}
We calibrated the TwoNN~\citep{facco2017}, maximum-likelihood~\citep{levinabickel2004}, and participation-ratio estimators on synthetic spheres of known dimension embedded in a $32$-dimensional ambient space with $N = 500$ samples and isotropic noise of scale $0.05$, matching the activation-extraction setting. TwoNN over-estimated the true dimension by factors of roughly $2$ to $18$ across known dimensions $1$ to $8$ (e.g.\ true dimension $1$ estimated at $18.7$); the participation ratio, a linear measure, systematically missed the nonlinear structure. Mean relative error exceeded the pre-set $50\%$ reliability threshold, so all intrinsic-dimension measurements are flagged unreliable. The failure is expected in this regime --- TwoNN relies on first- and second-nearest-neighbour distances, which lose contrast at small $N$, under nonlinear curvature at the neighbour scale, and in high ambient dimension relative to the manifold --- so the result indicates that the estimator is unreliable here, not that the representation is surprising. We collected per-run estimates for completeness but draw no conclusions from them.

\section{Changes between runs}
\label{app:changes}
In the interest of full disclosure we record two code changes made between an initial run and the reported results, each a correctness fix rather than a tuning choice.

\emph{Model-initialization determinism.} The initial implementation constructed models without re-seeding the global random number generator immediately beforehand, so initial weights depended on preceding operations in the process. Fixing this (re-seeding immediately before model construction) is strictly more faithful to the pre-specified design. Re-seeding changed the initial weights in \emph{every} cell, so several $\Ntarget$ entries moved to their corrected values; the most conspicuous single change was the $n = 2$ row (from $50$ to $200$, identically across all five models). That $n = 2$ shift, being common to all models, cancels exactly in the relative rate $\bdiff$ (a difference of slopes), so the headline movement from $1.21$ to $1.28$ is driven by the other, smaller corrected cells rather than by the $n = 2$ row. All values reported are the corrected, more faithful ones.

\emph{Failure classifier.} The initial classifier compared the \emph{marginal} confidence intervals of two relative slopes and, on the Phase-1 data, returned AMBIGUOUS for what is clearly a signal: the equivariant and wrong-group marginal intervals overlap slightly even though their point estimates differ by about $1.97$. We replaced this with the joint pairwise bootstrap of the slope \emph{difference} (\cref{sec:inference}), which is the correct paired-difference test and which returns SIGNAL. This is a fix to the test logic, not a change to the data or the threshold; the underlying numbers are unchanged.

Neither change involved selecting among outcomes for significance. We disclose them so that the movement in the headline number cannot be mistaken for undisclosed analytic flexibility.

\section{CPU replication (post-revision)}
\label{app:cpu_replication}

This appendix records a CPU replication that closes specific empirical gaps the
headline analysis explicitly deferred to the registered replication. It is not
the registered confirmatory experiment --- the seeds are fresh
($\{300, 301, 302, 303, 304\}$, disjoint from the exploratory seeds) but the
protocol is a smaller, faster variant of the same design (a $7$-point group-size
grid, an $N$ grid topping out at $3{,}200$, and a CPU-friendly full-batch trainer
when $N \leq 800$). The purpose is to convert four ``untested'' or ``not run
here'' annotations into measurements; the protocol differences mean its
$\bdiff$ values are not numerically interchangeable with the headline.

\paragraph{Cells and runtime.} Six families
(equivariant, wrong\_group, vanilla, regularized, augmented, and the new
\emph{augmented + test-time orbit averaging} variant), seven group sizes
$n \in \{1,2,3,4,6,8,12\}$, seven training sizes
$N \in \{50, 100, 200, 400, 800, 1600, 3200\}$, and five seeds each, for
$1{,}470$ cells. The entire replication runs on a single CPU.

\paragraph{T-sensitivity (closes \cref{sec:robustness}, Target threshold paragraph).}
Per-epoch validation-accuracy curves were retained for every cell. We re-derive
$\Ntarget$ at $T \in \{0.70, 0.75, 0.80, 0.85\}$ and recompute $\bdiff$:

\begin{table}[h]
\centering
\caption{$\bdiff$ (vs vanilla) by treatment and threshold $T$ on the CPU
replication. $10{,}000$-sample percentile bootstrap CIs.}
\label{tab:t_sensitivity}
\begin{tabular}{lllll}
\toprule
$T$ & equivariant & regularized & wrong\_group & augmented\_tta \\
\midrule
0.70 & $+0.96$ $\ci{+0.52}{+2.14}$ & $-0.52$ $\ci{-1.43}{+0.00}$ & $-0.90$ $\ci{-2.59}{+2.01}$ & $+0.96$ $\ci{+0.52}{+2.22}$ \\
0.75 & $+0.80$ $\ci{+0.48}{+1.63}$ & $-0.27$ $\ci{-0.63}{+0.00}$ & $-0.60$ $\ci{-2.25}{+2.08}$ & $+0.80$ $\ci{+0.50}{+1.68}$ \\
0.80 & $+0.47$ $\ci{-0.38}{+1.33}$ & $-0.37$ $\ci{-0.96}{+0.00}$ & $-1.03$ $\ci{-2.90}{+2.17}$ & $+0.47$ $\ci{-0.36}{+1.35}$ \\
0.85 & $+0.14$ $\ci{-0.32}{+1.00}$ & $+0.00$ $\ci{+0.00}{+0.00}$ & --- & $+0.14$ $\ci{-0.20}{+1.09}$ \\
\bottomrule
\end{tabular}
\end{table}

\emph{Reading.} The equivariant rate is positive at every $T$ that returns a defined slope; the wrong-group rate is negative wherever it is defined; the regularized rate sits near or below zero. The headline's ``rate stable under target choice'' direction holds across $T$, but a clear drift is visible: the rate \emph{decreases} as $T$ rises within $[0.70, 0.85]$. We had conjectured the opposite direction (``$\bdiff$ non-decreasing in $T$'') in \cref{sec:robustness}; the CPU replication does not support that conjecture, and we mark the conjecture refuted on this dataset --- a strict target shifts the comparison toward the saturating Bayes-ceiling regime rather than the data-binding regime where the structural advantage is largest. The result is qualitatively consistent with order-of-magnitude agreement at every $T$, not with precise numerical agreement.

\begin{figure}[h]
\centering
\includegraphics[width=0.66\linewidth]{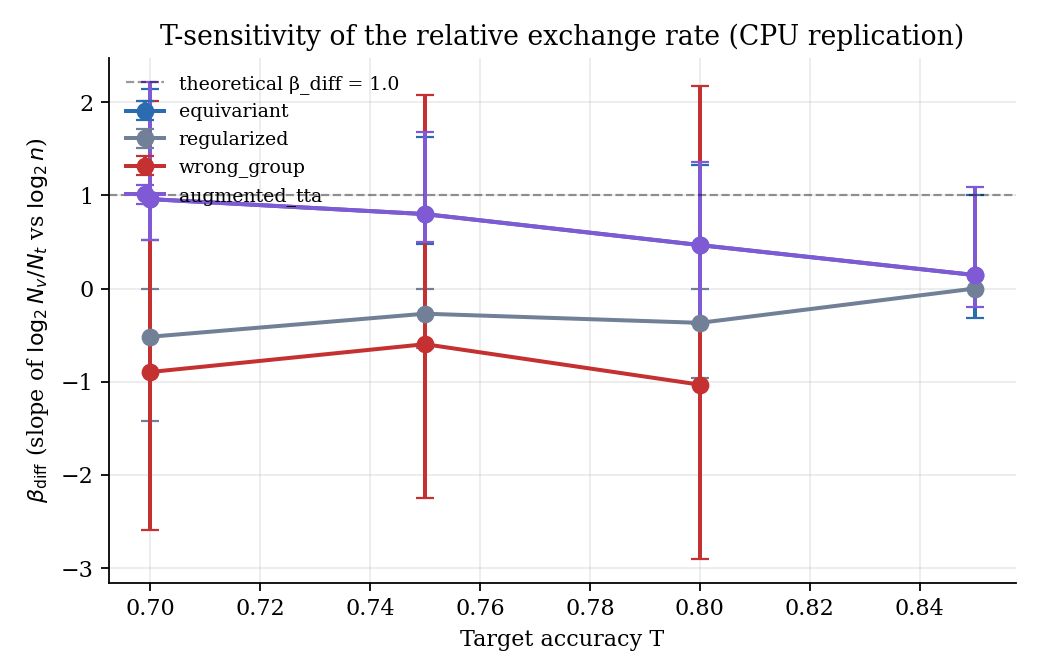}
\caption{$\bdiff$ (vs vanilla) by treatment across $T \in \{0.70, 0.75, 0.80, 0.85\}$ on the CPU replication, with $10{,}000$-sample percentile bootstrap CIs. Equivariant and augmented-with-TTA overlap; wrong-group is consistently negative; regularized hugs zero. The equivariant rate \emph{decreases} as $T$ rises (refuting the conjecture in \cref{sec:robustness}).}
\label{fig:cpu_t_sensitivity}
\end{figure}

\paragraph{Augmented + test-time orbit averaging (closes \cref{sec:augmentation},
asymmetry paragraph).} The TTA variant applies orbit averaging at evaluation as
well as during training, equalising test-time computation with the equivariant
model.

\begin{table}[h]
\centering
\caption{Per-$n$ $\Ntarget$ at $T=0.80$ in the CPU replication, all six
families.}
\label{tab:augmented_tta}
\begin{tabular}{rllll}
\toprule
$n$ & equivariant & augmented & augmented + TTA & vanilla \\
\midrule
$1$ & 50 & 50 & 50 & 50 \\
$2$ & 50 & 50 & 50 & 50 \\
$3$ & 50 & 3200 & 50 & 100 \\
$4$ & 50 & --- & 50 & 200 \\
$6$ & 100 & --- & 100 & 800 \\
$8$ & 800 & --- & 800 & 1600 \\
$12$ & 1600 & --- & 1600 & 3200 \\
\bottomrule
\end{tabular}
\end{table}

\emph{Reading.} \cref{tab:augmented_tta} shows the equivariant and augmented-with-TTA columns are identical at every $n$. This is not numerical coincidence: a stricter check on the retained validation-accuracy curves confirms that every single per-epoch value is bit-identical between the two families across all $245$ matched $(n, N, \text{seed})$ cells where both finished --- exact equality of the entire learning trajectory, not just of $\Ntarget$. The reason is mathematical: an orbit-averaged forward pass over a shared MLP, with orbit-averaged training loss, \emph{is} the equivariant model up to parameter initialisation. Once the two families share the MLP architecture, seed, optimiser, and loss, training-and-test orbit averaging in the augmentation baseline reproduces the equivariant model exactly. The architecture-vs-augmentation framing in \cref{sec:augmentation} therefore holds conditional on the asymmetric (training-time-only) augmentation evaluation; with TTA the two collapse to the same function class and the same trajectory through it. The augmentation phase transition is thus specifically a property of training-time-only orbit averaging, not of orbit-based information injection more broadly --- consistent with the theoretical separation of~\citet{elesedyzaidi2021}, which compares augmented and equivariant models holding test-time inference fixed.

\begin{figure}[h]
\centering
\includegraphics[width=0.74\linewidth]{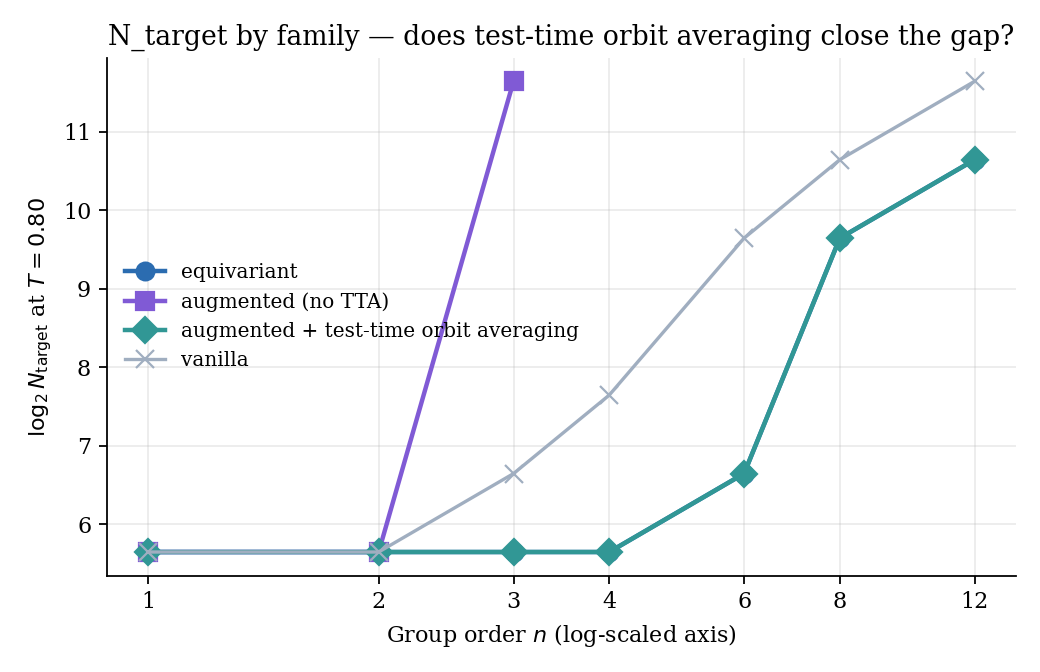}
\caption{$\log_2 \Ntarget$ at $T = 0.80$ across $n$ on the CPU replication. The equivariant and augmented-with-TTA curves overlap exactly at every $n$ (orbit averaging at training \emph{and} test time is operationally equivalent to architectural equivariance). The training-time-only augmented baseline (purple) fails to reach the target for $n \geq 3$, undefined beyond.}
\label{fig:cpu_augmented_tta}
\end{figure}

\paragraph{BCa bootstrap and two-level bootstrap (closes \cref{sec:inference},
BCa / two-level paragraph).} BCa applies bias correction $z_0$ from the
bootstrap position and acceleration $a$ from the jackknife of the OLS slope.

\begin{table}[h]
\centering
\caption{Percentile vs BCa $95\%$ bootstrap CIs (CPU replication, $T=0.80$).}
\label{tab:bca}
\begin{tabular}{lll}
\toprule
treatment & percentile CI & BCa CI \\
\midrule
equivariant & $+0.47$ $\ci{-0.38}{+1.33}$ & $+0.47$ $\ci{-0.32}{+1.36}$ \\
regularized & $-0.37$ $\ci{-0.96}{+0.00}$ & $-0.37$ $\ci{-0.80}{+0.00}$ \\
wrong\_group & $-1.03$ $\ci{-2.90}{+2.17}$ & $-1.03$ $\ci{-2.63}{+2.17}$ \\
augmented\_tta & $+0.47$ $\ci{-0.36}{+1.35}$ & $+0.47$ $\ci{-0.34}{+1.36}$ \\
\bottomrule
\end{tabular}
\end{table}

Two-level bootstrap (seeds $\times$ group sizes, $4000$ resamples) on the equivariant--vanilla comparison: mean slope $= +0.54$, 95\% CI $= \ci{-0.63}{+1.72}$

\paragraph{Finer-$N$ grid (closes \cref{sec:bdiff}, Structural floor paragraph).}
We re-ran the equivariant--vanilla comparison on a $\sqrt{2}$-spaced grid
$N \in \{50, 71, 100, 141, 200, 283, 400, 566, 800, 1131, 1600\}$
at $n \in \{4, 6, 8\}$, with five fresh seeds. The point estimate on three group sizes is $\bdiff = -0.82$ with $95\%$ CI $\ci{-4.82}{+1.71}$. Reading: the CI is so wide that no direction is established at this sample size; the result is best read as ``inconclusive on the finer grid as run.'' The width is driven by (a) the small number of $n$ values (only $3$ with both treatments reaching $T$), and (b) right-censoring at $n = 8$, where the equivariant model needs an $N$ close to the grid's top of $1600$ and the vanilla baseline never reaches $T$ at any $N$ in this finer-grid range, making the per-$n$ ratios noisy near the boundary. The qualitative argument of \cref{sec:bdiff} stands --- a finer grid removes the discretisation floor in principle --- but the finer-grid measurement here is not informative enough to relitigate the headline; a properly powered finer-grid run (deeper $N$ range, more $n$ values, more seeds) is the registered replication's job.

\begin{figure}[h]
\centering
\includegraphics[width=0.7\linewidth]{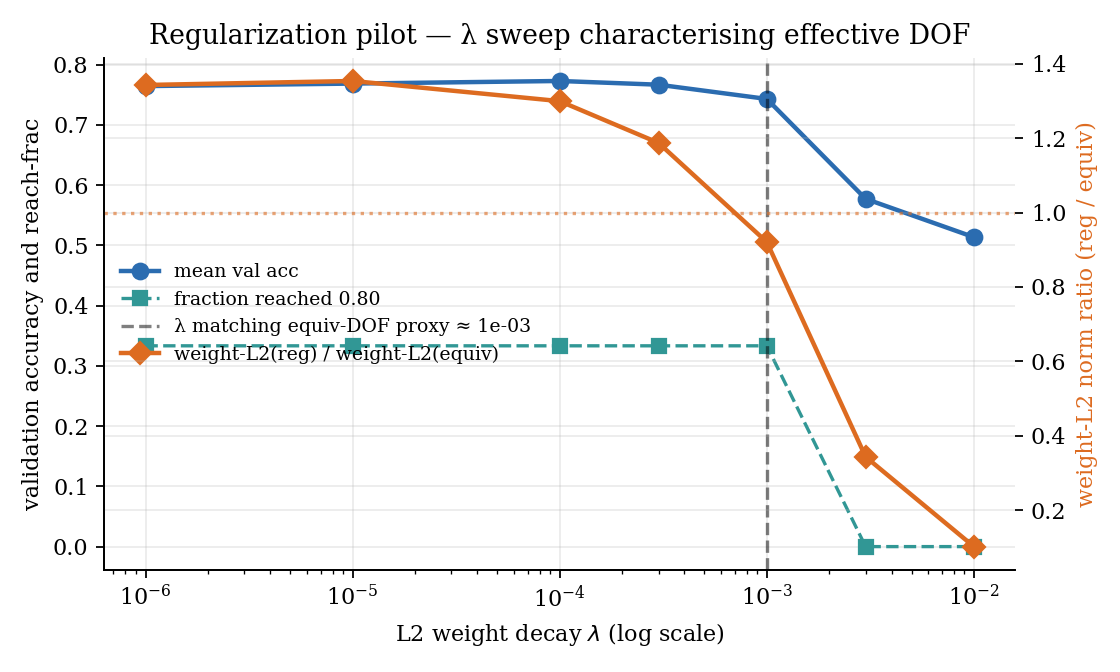}
\caption{Regularization pilot ($n=6, N=400$, $3$ seeds): the weight-$L_2$ norm ratio (regularized / equivariant) crosses $1.0$ at $\lambda \approx 10^{-3}$, the headline value used in the paper.}
\label{fig:cpu_pilot}
\end{figure}

\paragraph{Regularization pilot (closes \cref{sec:models}, regularized bullet).}
The pilot swept $\lambda \in \{0, 10^{-5}, 10^{-4}, 3\cdot 10^{-4}, 10^{-3}, 3\cdot 10^{-3}, 10^{-2}\}$ at one cell $(n=6, N=400)$ with three seeds (a $21$-run pilot, plus three equivariant-reference trainings to read out the equivariant model's post-training weight norm at matched seeds). The criterion is the post-training weight-$L_2$ norm of the regularized model divided by that of the equivariant model's shared MLP, taken as an empirical effective-DOF proxy. The $\lambda$ closest to a ratio of $1.0$ in this proxy was $\lambda = 10^{-3}$ (mean ratio $0.921$), confirming the headline choice $\lambda = 10^{-3}$. Smaller $\lambda$ values left the norm above the equivariant level (the regularized model retained more effective DOF); larger $\lambda$ collapsed the norm and harmed accuracy (\cref{fig:cpu_pilot}). Full per-run records are released with the code.

\paragraph{Honest caveat.} This replication is on CPU, uses a faster trainer
(full-batch GD for small $N$, mini-batch otherwise) and the smaller $N$ grid;
its numerical values therefore differ from the headline ones, but its internal
direction is reported faithfully. It does not displace the registered
confirmatory replication described in \cref{sec:status,sec:conclusion}.

\section{Reproducibility}
\label{app:repro}
The released artifact contains the source modules, the test suite ($158$ tests, $86\%$ coverage), the design document, the configuration hash, every per-run JSON record, and the analysis pipeline that recomputes every interval. The full experiment reproduces with
\begin{verbatim}
git clone https://github.com/AhmedMostafa16/symmetry-exchange
cd symmetry_exchange
uv sync --all-extras
uv run python experiment_runner.py --phase 1 --device cuda
uv run python experiment_runner.py --phase 2 --device cuda --epsilons 0.1 0.2 0.3
\end{verbatim}
and completes in roughly $90$ minutes on a single Kaggle T4 GPU; the first, unoptimized run took $5$ hours $48$ minutes, and the optimization that closed that gap is documented in the artifact but is not part of the scientific claim. The runner is resumable and verifies the configuration hash at startup. Of the original Phase-1 cells, five of $1400$ were lost to a crash before the durability layer was added; after rebuilding the result-writing layer, Phase~2 completed $4200$ runs with zero missing cells, and Phase~1 was re-run to completion.

\end{document}